\documentclass[sigconf,natbib=true,anonymous=false]{acmart}

\usepackage{multirow}
\usepackage{graphicx}
\usepackage[ruled,noend]{algorithm2e}
\usepackage{subfigure}
\usepackage[misc]{ifsym}
\usepackage{subfigure}
\usepackage{float}

\AtBeginDocument{%
  \providecommand\BibTeX{{%
    \normalfont B\kern-0.5em{\scshape i\kern-0.25em b}\kern-0.8em\TeX}}}

\setcopyright{acmcopyright}
\copyrightyear{2022}
\acmYear{2022}

\acmConference[MM '22] {Proceedings of the 30th ACM International Conference on Multimedia }{October 10--14, 2022}{Lisbon, Portugal.}
\acmBooktitle{Proceedings of the 30th ACM International Conference on Multimedia (MM '22), October 10--14, 2022, Lisbon, Portugal}
\acmPrice{15.00}
\acmISBN{978-1-4503-9203-7/22/10}
\acmDOI{10.1145/XXXXXX.XXXXXX}

\begin{document}

\title{Adaptive Structural Similarity Preserving for Unsupervised Cross Modal Hashing}



\author{Liang Li}
\affiliation{%
\institution{School of Computer Science\\ Fudan University}
\country{China}
}
\email{lil19@fudan.edu.cn}

\author{Baihua Zheng}
\affiliation{%
  \institution{School of Computing and Information Systems, Singapore Management University}
  \country{Singapore}
}
\email{bhzheng@smu.edu.sg}

\author{Weiwei Sun}
\affiliation{%
  \institution{School of Computer Science\\Fudan University}
  \country{China}
}
\email{wwsun@fudan.edu.cn}

\renewcommand{\shortauthors}{Trovato and Tobin, et al.}

\begin{abstract}
Cross-modal hashing is an important approach for multimodal data management and application. 
Existing unsupervised cross-modal hashing algorithms mainly rely on data features in pre-trained models to mine their similarity relationships.
However, their optimization objectives are based on the static metric between the original uni-modal features, without further exploring data correlations during the training. In addition, most of them mainly focus on association mining and alignment among pairwise instances in continuous space but ignore the latent structural correlations contained in the semantic hashing space. 
%
In this paper, we propose an unsupervised hash learning framework, namely \emph{Adaptive Structural Similarity Preservation Hashing (ASSPH)}, to solve the above problems. Firstly, we propose an adaptive learning scheme, with limited data and training batches, to enrich semantic correlations of unlabeled instances during the training process 
and meanwhile to ensure a smooth convergence of the training process.
Secondly, we present an asymmetric structural semantic representation learning scheme. We introduce structural semantic metrics based on graph adjacency relations during the semantic reconstruction and correlation mining stage and meanwhile align the structure semantics in the hash space with an asymmetric binary optimization process. Finally, we conduct extensive experiments to validate the enhancements of our work in comparison with existing works.
\end{abstract}




\maketitle

\section{Introduction}

Hash coding has become an essential method for similarity queries due to its computational efficiency and storage cost advantages. A series of hash representation learning methods~\cite{wang2017survey,cao2017hashnet,jiang2017deep} developed from local similarity hashing further improve the hashing retrieval efficiency. However, hash learning needs to consider additional factors for multi-modal data.
In cross-modal hashing, instances from different modalities
need to be compared and ranked together, which means the hamming distance of instances from different modalities can reflect the shared semantic relationship.

Thanks to the rapid development of representation learning, current cross-modal hashing methods often perform fine-tuning on the backbone encoding models with a joint-optimization alignment loss and a binary optimization strategy to obtain the final hashing model. High-quality encoding offered by the pre-trained models predominantly improves the cross-modal retrieval performance, especially the unsupervised performance.
In recent years, unsupervised models, such as UnifiedVSE~\cite{wu2019unified}, VL-BERT~\cite{su2019vl}, and VLP~\cite{zhou2020unified}, have made breakthroughs on cross-modal retrieval tasks. In this paper, our hash-learning targets at a more general unsupervised cross-modal scenario~\cite{su2019deep,liu2020joint,zhang2021high}, where hash models are learned to map the multimodal semantic feature vectors to a binary space such that the hash-learning can improve the performance of multimedia data management and retrieval in a broad manner. 
Meanwhile, our research focuses on the representative task of image-text cross-modal retrieval. 
However, as our method is not limited to feature extraction backbones, it can be extended to other cross-modal retrieval between/among other modalities. 

According to optimization objectives, existing unsupervised cross-modal hashing methods can be clustered into alignment methods~\cite{zhang2018unsupervised} and reconstruction methods~\cite{su2019deep,hu2020creating,yu2021deep}.
Both classes of methods lack reliable cross-modal similarity as the supervision during the training~\cite{zhang2021high}. Although some in-depth studies~\cite{yu2021deep,zhang2021high} have been conducted in constructing more accurate cross-modal uniform metrics, they ignore the fact that the final goal of our hashing study is to obtain a plausible cross-modal metric, which is difficult to achieve based on a simple combination of raw multi-modal similarity distribution.

To exploit the inter- and intra-modal semantic relationships, existing unsupervised hashing methods have widely adopted graph-based paradigms to regularize similarity relationships. However, existing methods often suffer from the ``static graph'' problem~\cite{shen2020auto}. More specifically, they often employ features from original data or pre-trained models to build the explicit precomputed graphs. However, those pre-defined graphs cannot be adaptively learned during the training to better express the semantic structures. 
Different from the label-based reliable static association graph in the supervised methods~\cite{chen2021local,liu2021graph}, the ''static graph'' constructed based on the original data measurement in the unsupervised methods~\cite{liu2020joint,yu2021deep} actually introduces the bias in the original feature measurement. 

As there is no reliable static optimization objective in unsupervised cross-modal hash learning, the difficulty in imporving the unsupervised performance lies in how to smoothly refine the optimization objective and reduce the prior noise in the training process, thereby improving the performance. It has been confirmed in existing work~\cite{caron2018deep} that a deep representation learning model with prior knowledge can be trained on middle results and iteratively converge. In recent research, the mainstream unsupervised research~\cite{he2020momentum} provides excellent examples through the optimization of memory banks and various variants. However, the memory-bank-based methods require massive training data as the basis, which is difficult to obtain for general application scenarios. 

We aim at optimizing the optimization goal (``static graph'') in the learning process to mine more information under unsupervised conditions. In order to ensure a smooth convergence of training, we do not directly adopt the continuous similarity between intermediate results, but propose a two-stage correlational relationships mining strategy to get more reliable prior information. This relationship also serves as additional pseudo-supervision information to resolve the prior noise in the original static graph. In brief, in this paper, we propose a novel adaptive hash learning framework based on adaptive structural similarity preservation to overcome the above problem from two aspects. 
Firstly, we design an adaptive correlation mining scheme to provide reliable positive samples as supervision while being able to smoothly update the correlational relationship set during the training to enrich the semantics correlations. 
Secondly, we introduce a structural semantic representation and an asymmetric binary code learning scheme together to preserve the semantic similarity in the final hash space. 

In summary, this paper makes a three-fold contribution.
First, we propose a novel adaptive cross-modal correlation mining and structural semantic maintenance strategy. The optimization target is updated indirectly by adaptively expanding the semantic-similar neighborhood to solve the ``static graph'' problem and maintain a smooth convergence of the optimization process. 
Though traditional unsupervised learning methods are strictly limited by static constraints, our method does not rely on expensive calculation and data resources in contrastive learning and can be widely adopted in hash retrieval optimization.
%
%
Second, we introduce a novel cross-modal semantic preservation framework as the backbone of our hash learning to bridge the modality gap. Specifically, we learn common hash representations through multi-level structural semantic consistency constraints, which serve as the basis of adaptive learning. In addition, we design an asymmetric similarity-preserving binary optimization algorithm to reduce the information loss after binarization.
%
%
Last but not least, we perform extensive experimental studies on two publicly available datasets, NUSWIDE and MIRFlickr-25K, and our approach is able to significantly outperform state-of-the-art methods proposed in the last three years.

\begin{figure*}[t]
  \centering
  \includegraphics[width=0.9\textwidth]{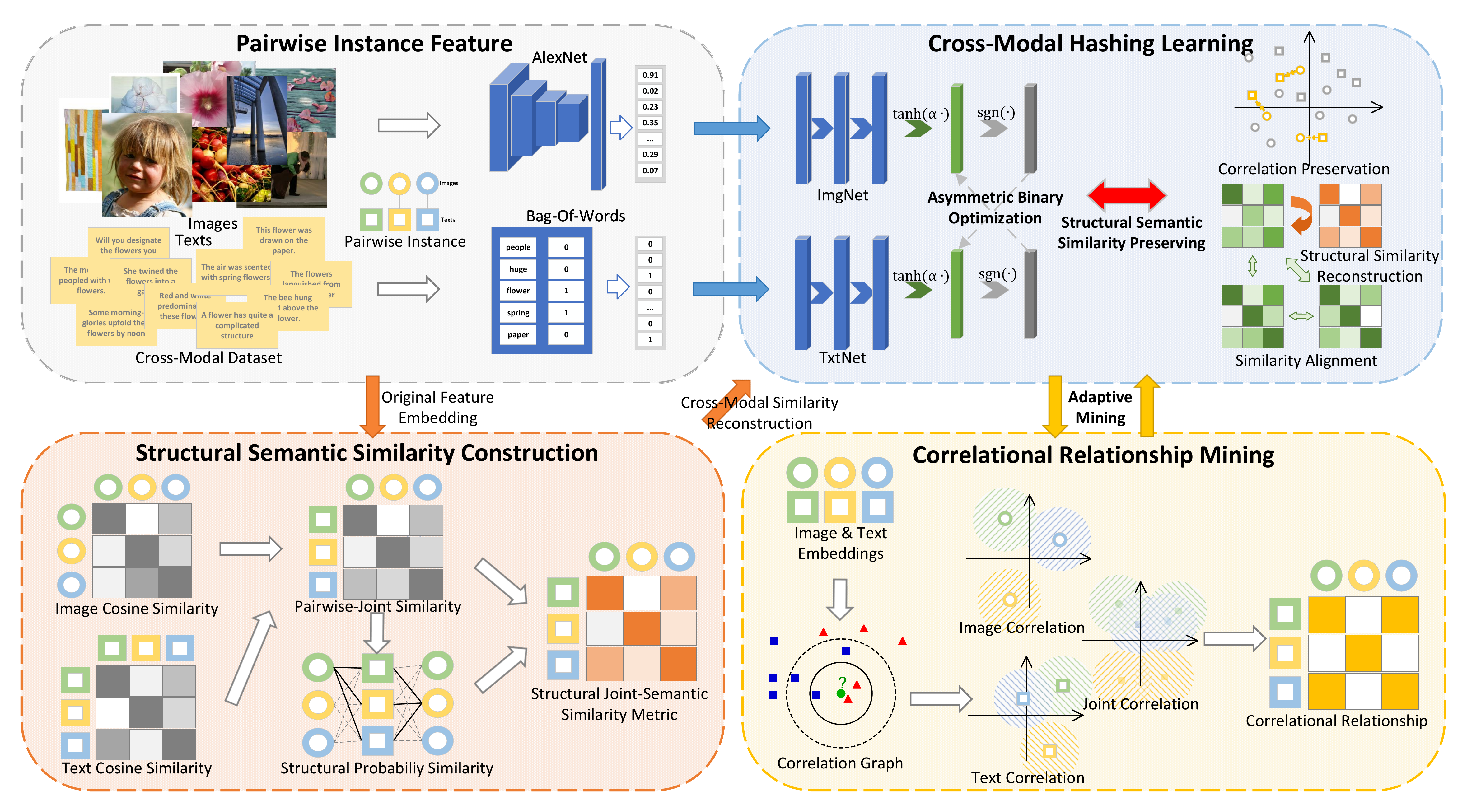}
  \vspace{-0.1in}
  \caption{An overview of our Adaptive Structural Similarity Preserving Hash (ASSPH) model.} 
  \label{fig:assph}
  \vspace{-0.2in}
\end{figure*}

\section{Related Works}

Hashing learning~\cite{wang2015learning,wang2017survey} in uni-modal and cross-modal datasets aims to learn a hashing function to map the original data to a binary hashing space while the hashing metric is expected to preserve the semantic similarities between original multimedia data. 
For cross-modal hashing, the additional modality gives new opportunities to describe and dig up the correlation between/among modalities, and thus, the contrastive-based methods can perform well. However, due to the distribution difference in various data modalities, the structural-semantic difference, called the \emph{semantic gap}, impedes the construction of semantic mapping. As labels play an important role in existing cross-modal learning to determine whether two instances are similar, related works in the past decade can be categorized into the following two classes, i.e., \emph{supervised hashing} and \emph{unsupervised hashing}.

\vspace{0.03in}
\noindent
\textbf{Supervised Cross-Modal Hashing.}
%
Early supervised cross-modal hashing methods, e.g., Semantic Correlation Maximization (SCM)~\cite{zhang2014large} and Semantics-Preserving Hashing (SePH)~\cite{lin2015semantics}, perform the reconstruction via different approaches, including semantic similarity matrix preserving and KL-divergence based correlation measurement. 
Recent works~\cite{wang2017adversarial,jiang2017deep,hu2018deep,li2018self} start using the deep neural network and semantic embeddings provided by pre-trained models. 
A subsequent series of works~\cite{liu2018fast, chen2019two, wang2020label,liu2019mtfh,meng2020asymmetric, yang2021rethinking} explore the relationships of semantic tags and latent semantics with different conditions and improve the hashing performance.
Supervised methods achieve good performance by using labels as supervision information to align different modalities; however, due to the high cost of labeling, they cannot be widely used. On the contrary, unsupervised methods, to be reviewed next, are not limited by labels and thus can be applied to a wider range of applications.


\vspace{0.02in}
\noindent
\textbf{Unsupervised Cross-Modal Hashing.}
%
Unsupervised cross-modal hashing methods~\cite{hardoon2004canonical,long2016composite,weiss2008spectral,ding2016large} aim to maximize the correlation between different modals or to preserve the inter-modal and intra-modal similarity, like Canonical Correlation Analysis (CCA)~\cite{hardoon2004canonical} and Composite Correlation Quantization (CCQ)~\cite{long2016composite}. With the help of deep representation learning, some recent methods~\cite{lee2013pseudo,caron2018deep} can describe inter-modal correlations more precisely and perform well on semantic alignment.



According to their optimization objectives, recent unsupervised hashing methods can be broadly divided into alignment methods and reconstruction methods. 
To overcome the lack of semantic supervision under unsupervised conditions, the first category usually utilizes some data-mining methods to describe the correlational relationship (pairwise or triplet relationship), a measurement of the similarity or semantic difference between two cross-modal instances~\cite{hu2017pseudo,zhu2016unsupervised,he2017unsupervised,wu2018unsupervised,zhang2018unsupervised,wang2020set,wang2021cluster}. Some contrastive-learning-based frameworks are introduced to learn the hash representation. Some works~\cite{gu2018look,li2019coupled} even try to use the generative models~\cite{mirza2014conditional} to describe the distribution of each modality and then make an alignment while introducing additional overhead during the hash coding.

The second category adopts the semantic reconstruction~\cite{su2019deep, yang2020deep, liu2020joint, yu2021deep,zhang2021high}, which considers the similarity as the exact distance value in the hash space instead of relevance similarity or probability. While keeping the semantic metric in the hash space directly, they are all constrained by the noise arising in the feature extraction. Some works~\cite{wang2020set,zhang2021high} use a two-stage optimization and independent codebook to update the similarity guideline simultaneously, e.g.,  UKD~\cite{hu2020creating} proposes a two-stage data distillation framework to the cross-modal hashing learning and demonstrates the effectiveness of the cross-modal hidden embeddings to optimize the following cross-modal hashing network further.

\section{Problem Formulation}

This paper focuses on cross-modal retrieval between image and text modalities. Let $O_i = \{v_{i},t_{i}\}$, $i \in \{1, 2, \cdots, M\}$ indicate an $i^{th}$ cross-modal instance, $M$ represent the total number of instances, $v_{i}\in \mathbb{R}^{d_{v}}$/ $t_{i}\in \mathbb{R}^{d_{t}}$ refer to the visual/text feature in the pairwise instance, and $d_{v}$/$d_{t}$ represents the dimension of the image/text features. 
   
Given the cross-modal set \begin{math}O\end{math} with the corresponding image feature set \begin{math}I=\{v_{i}\}_{i=1}^{M}\end{math} and text feature set \begin{math}T=\{t_{i}\}_{i=1}^{M}\end{math}, we want to learn hash functions \begin{math}f_{I}(v,\theta ^{I})\end{math} and \begin{math}f_{T}(t,\theta ^{T})\end{math} to generate hash code \begin{math}B^{I}, B^{T}\in \{-1,+1\}_{}^{M\times K}\end{math} 
for image and text modalities respectively. $K$ is the length of the hash code, $\theta ^{I}$ and $\theta ^{T}$ are the parameters of ImgNet and TxtNet for hash code to be optimized. In addition, the similarity between hash codes $b_{i}^{I}$ and $b_{j}^{T}$, i.e., their Hamming distance $D(b_{i}^{I},b_{j}^{T})=\frac{1}{2}(K-\left \langle b_{i}^{I},b_{j}^{T} \right \rangle)$, is able to reflect the semantic similarity between original instances $v_{i}$ and $t_{j}$. 

\section{Proposed Approach}

Despite recent significant breakthroughs in unsupervised work based on contrast learning~\cite{he2020momentum,chen2020simple,grill2020bootstrap,chen2021exploring}, it is still challenging to apply this learning paradigm in many scenarios with limited data and computational resources. The main problem is that it is difficult to obtain enough stable intermidiate representations as supervision to optimize the model during the learning process smoothly. Existing unsupervised cross-modal hashing methods use the metric relationship within the original features of the data as the optimization objective directly or use data distillation~\cite{hu2020creating} to divide the training process into multiple stages. However, these approaches amplify the interference of noise in the original features, which limits the performance of unsupervised hashing.

In this paper, we propose \emph{Adaptive Structural Similarity Preserving Hash (ASSPH)} model, as a solution to the problem formulated above. ASSPH designs a comprehensive 
structured semantic retention strategy with a new positive sample expansion technique to stablize the training process while smoothening the updates of the optimization target to enrich the cross-modal semantic~\cite{chang2017deep,caron2018deep}. In addition, ASSPH introduces an asymmetric deep hash learning method to further reduce the quantization loss and to preserve the structural semantic in the hashing space to improve the retrieval performance. ASSPH consists of four major components, as shown in Figure~\ref{fig:assph}, and Algorithm~\ref{alg:TEmax} provides the pseudocode of the proposed hash learning method. In the rest of this section, we will detail the four components. 

\subsection{Feature Extraction}

Feature extraction is an independent component in ASSPH. We follow the setup of existing work and use AlexNet and Bag-of-Words for extracting image feature $F^{I}$ and text feature $F^{T}$ respectively. Switching to a more efficient model like VGG or ResNet does improve the final result, but it is not the main focus of this work. ASSPH supports different feature extraction models in order to support a wider range of application scenarios.

\subsection{Structural Semantic Similarity Construction}

Once the original semantic features $F^{I}$ and $F^{T}$ of the cross-modal training set $O=\{(v_{i}, t_{i})\}_{1}^{M}$ are obtained, ASSPH constructs their structural similarities. It first transforms the original semantic metric into a similarity probability based on the metric distribution and practical application implications; it then defines the joint-modal similarity and structural semantic similarity between pairs of data respectively; it finally combines these two similarities together to form the structural semantic similarity metric.

%
To be more specific, we firstly follow the previous works~\cite{hu2018deep} to calculate the pair-wise modality-specific cosine metrics within different modals, $S^{*}=\{s_{ij}^{*}=\cos(f_{i}^{*},f_{j}^{*})\}_{i,j=1}^{M}$ to capture pair-wise semantic similarities between instances within a single modality. Note, in this paper, $* \in \{I, T\}$ represents either image or text modality.
As we find that the distribution shape of $S^{*}$ is approximately a skewed normal distribution, we map $S^{*}$ from $[-1,1]$ to the interval $[0,1]$, denoting it as $S^{*}_p=2S^{*}-1$. $S^{*}_p$ is regarded as the probability that can describe whether two instances in a single modality (i.e., $v_i$ and $v_j$ in image modality or $t_i$ and $t_j$ in text modality) are semantically similar. We then calculate its cross-modal similarity $S_{fusion}$ under the independence condition to subtract the common part after adding the probabilities.
\begin{equation}
S_{fusion} = S_{p}^{I} + S_{p}^{T} - S_{p}^{I} \cdot S_{p}^{T}
\label{equ:s_fusion}
\end{equation}

To define the structural semantic, existing works~\cite{su2019deep,liu2020joint} directly compare the fusion similarity of neighbor distance distributions between instances as the structural similarity. However, we observe that when the fusion similarity value is small, its accuracy as a metric is correspondingly reduced. Take the fusion similarity constructed on MIRFlickr as an example. MAP@50 is around 0.9 when MAP@all decreases to about 0.75. To tackle this issue, we incorporate two optimizations. Firstly, we keep only the top $K_{S}$ nearest distances based on fusion similarity $S_{fusion}$, but remove the rest of distances that are much noisier and less useful. In our implementation, $NN_{K_S}(i)$ w.r.t. an instance $i$ captures its top-$K_{S}$ nearest neighbors. 
Secondly, we reweight the distances between the remaining instances based on their original cosine distance to reduce the impact of uneven density.
That is to say, if $j\in NN_{K_S}(i)$,
$\hat{S}(i,j)=\frac{S_{fusion}(i,j)}{ {\textstyle \sum_{k \in NN_{K_S}(i)}^{} S_{fusion}(i,k)} }$; otherwise, $\hat{S}(i,j)=0$.
The structural similarity is defined in Eq.~(\ref{equ:s_str}). 
Here, $K_{S}$ is used again to rebalance data range; $\hat{S}^{'}$ refers to the matrix transpose of $\hat{S}$, and we will follow this notation later. 
%

\begin{equation}
S_{structure} = K_{S} \hat{S} \times \hat{S}^{'} \label{equ:s_str}
\end{equation}

Based on the above definition, we define the structural semantic similarity $S$ by considering both cross-modal similarity and structural similarity as a multi-level semantic metric to fully represent the structural semantic information. We finally map $S$ back to the range of the cosine metric, i.e. $S=2S^{'}-1$. 
\begin{equation}
S^{'}=(1-\gamma)S_{fusion} + \gamma S_{structure}, \gamma \in [0,1]
\label{S_CAL}
\end{equation}

\vspace{-0.03in}
\subsection{Correlational Relationship Mining}

Our ASSPH model proposes to construct a set of correlational relationships as the positive samples to further guide the pair-wise cross-modal hashing alignment. In addition to correcting bias in static structural semantic similarity, we introduce the correlational relationships because of the following considerations. 
On the one hand, limited by the similarity distribution between the original data, the existing static similarity measures~\cite{su2019deep,liu2020joint,yu2021deep} tend to have poor local consistency. That is, most of the measurement results are in the insensitive range, which makes it challenging to effectively map correlated instances into adjacent hash codes.
On the other hand, adjusting the whole continuous similarity metric space directly can have uncontrollable effects on the training process and lead to training collapse, especially when the training data is not large enough.  
Inspired by DeepCluster~\cite{caron2018deep}, we try to solve this problem by revising the discrete constraints. We introduce a new discrete training target here, which can use the knowledge learned by the model during each training step to further strengthen the local consistency of the hash representation and ensure that the training process can eventually converge smoothly. 

%
%
To be more specific, we define a strict correlation discrimination scheme based on the common adjacency relations in inter- and intra- modals. We initialize the correlated instance set by considering their original semantic features, and then adaptively expand the set based on the currently learned cross-modal hidden-layer representations during the model training process.
%

\vspace{0.03in}
\noindent
\textbf{Initialization of the Correlational Relationship.}
We introduce a two-stage positive sample expansion strategy to initialize the the binary-valued correlational relationship set defined as $R=\{r_{ij} \in \{0, 1\}\}_{i,j=1}^{M}$. In the first stage, we perform proximity instance mining based on $S^{I}$ and $S^{T}$ (the pair-wise modality-specific cosine metrics introduced previously) respectively. To be more specific, for $j\in NN_{K_R}(i)$, 
$R_{1}^{*}[i,j]=1$; otherwise, $R_{1}^{*}[i,j]=0$. 
%
Note $K_{R}$ controls the range of neighborhood. To preserve the semantic consistency of proximity instances, $K_{R}$ will be much smaller than $K_{S}$ and it will be dynamically expanded in adaptive similarity learning to be detailed next. 

However, KNN-based results generated in the above stage still consist of many errors, e.g., the neighboring instances of some marginal data or outliers often have different semantics. Therefore, we further define the second-order proximity relation $R_{2}^{*}$ to measure the similarity between intances by their shared neighbors. If $R_{1}^{*} \times R_{1}^{*'} \ge \tau $,  $R_{2}^{*}=1$; otherwise, $R_{2}^{*}=0$. 
%
Here, $\tau$ is a similarity threshold, and it is set to 1 in practice to ensure better results. 
Similarly, we can define the cross-modal correlation $R_{2}^{cross}$. If max($R_{1}^{I} \times R_{1}^{T'},R_{1}^{T} \times R_{1}^{I'}) \ge \tau$,
$R_{2}^{cross}=1$; otherwise, $R_{2}^{cross}=0$. 
Finally, Eq.~(\ref{R_CAL}) defines the binary-valued correlational relationship $R$.
\begin{equation}
R=R_{2}^{I} \cup R_{2}^{T} \cup R_{2}^{cross} \label{R_CAL}
\end{equation}

\vspace{0.03in}
\noindent
\textbf{Adaptive Similarity Learning.} Adaptive learning aims to keep the optimization target consistent~\cite{he2020momentum} to avoid training collapse due to convergence to a trivial solution~\cite{grill2020bootstrap}. To maintain stable iterations during the training process under the conditions of limited training batches and small datasets, we adopt a clear and effective expansion strategy based on the correlational relationship defined above. 

For each epoch online, we update the correlational relationship. After the $e^{th}$ training round, we can obtain hidden embeddings $H$ containing cross-modal semantics from the latest hash functions $ImgNet(F_{I},\theta_{e}^{I})$ and $TxtNet(F_{T},\theta_{e}^{T})$, i.e., $H_{i}^{I}=ImgNet_{e}(F^{I},\theta_{e}^{I})$ and $H_{i}^{T}=TxtNet_{e}(F^{T},\theta_{e}^{T})$.
Then, we replace the original features with the hidden features of the current hash layer and use the two-stage correlation mining strategy presented above to obtain new cross-modal relationships $R_{e}$. Finally, we union it with the current $R_{e-1}$, as stated in Eq.~(\ref{R_UPDATE}).
\begin{equation}
R_{e}=R_{e-1} \cup R(H^{I}, H^{T}) \label{R_UPDATE}
\end{equation}

To ensure convergence, we adopt a lower learning rate while strictly limiting the size of $K_{R}$ to control the reliability of the positive correlations. Subsequent experiments demonstrate that our correlation mining process further enriches the semantic correlation distribution while maintaining convergence. Due to structural semantic construction, our method requires $O(M^2)$ complexity for training, where $M$ represents the training set size. We argue that it is acceptable in most cases because of limited training set size.

\subsection{Cross-Modal Hashing Learning}

The key point of hash learning is to ensure that the distance relationship between hash codes can reflect the semantic similarity of original data instances. However, it is challenging to optimize binary representations in gradient backpropagation. Therefore, in our cross-modal hash learning module, we first design a set of cross-modal alignment loss functions to optimize the continuous hash hidden layer activated by the $tanh()$ function, and then introduce an asymmetric binary optimization approach to improve the semantic expression ability of the binary hashing representation.


\vspace{0.03in}
\noindent
\textbf{Structural Semantic Similarity Preservation.} We adopt two independent networks $ImgNet(F^{I}$, $\theta ^{I})$ and $TxtNet(F^{T}, \theta ^{T})$ to get the continuous cross-modal representations. Take the hidden features $H^{I}$ from the $ImgNet()$ and $H^{T}$ from the $TxtNet()$ as input, we design a multi-task learning paradigm to jointly learn the hash code to maintain the semantics within the original cross-modal data. The optimization task consists of three parts: structural semantic reconstruction, correlational relationship preservation, and semantic alignment. 

First, we follow the similarity preservation method in existing unsupervised learning approaches. Existing cross-modal unsupervised methods can be classified into contrastive learning, autoencoders, and similarity preservation, according to the category of the loss function used. Among them, similarity preservation has the lowest dependence on training batch size, which benefits scenarios with limited computational resources. Therefore, we first define a similarity preserving loss in Eq.~(\ref{equ:L_sr}). That is, both intra-modal and inter-modal metric results should be consistent with a pre-defined structural similarity metric. 
\begin{equation}
\begin{split}
\mathcal{L}_{sr}(H^{I}, H^{T})&=\left \| S-cos(H^{I},H^{T}) \right \|_{F}^{2}  + \left \| S-cos(H^{I},H^{I}) \right \|_{F}^{2} \\ &+
\left \| S-cos(H^{T},H^{T}) \right \|_{F}^{2} \label{equ:L_sr}
\end{split}
\end{equation}

Second, we want to achieve symmetric consistency,  
i.e., both intra and inter-modal similarities are consistent. Therefore, we construct a semantic alignment loss in Eq.~(\ref{LOSS2}) to ensure that both the intra-modality and inter-modality metrics are consistent. 
\begin{equation}
\begin{aligned}
\label{LOSS2}
\mathcal{L}_{sa}(H^{I}, H^{T})&=\left \| cos(H^{I},H^{I})-cos(H^{T},H^{T}) \right \|_{F}^{2}  \\& + \left \| cos(H^{I},H^{T}) - cos(H^{I},H^{I}) \right \|_{F}^{2} \\&+
\left \| cos(H^{I},H^{T})-cos(H^{T},H^{T}) \right \|_{F}^{2} 
\end{aligned}
\end{equation}

As the above-mentioned two optimization goals are both
based on static graph and hence will be affected by the prior deviation of the static graph, we introduce the set of correlational relationships $R$ 
to rectify the training process. That is, the distance of related instances shall be as close as possible. Meanwhile, considering the massive difference in the number of positive and negative instance pairs, we introduce a re-balance coefficient $\beta$ to optimize the performance. The semantic consistency loss is defined as follows:
%
\begin{equation}
\begin{split}
\label{LOSS3}
\mathcal{L}_{cp}(H^{I}, H^{T})=\left \| cos(H^{I},H^{T})\cdot R - \beta R \right \|_{F}^{2} 
\end{split}
\end{equation}

Eq.~(\ref{equ:L}) defines the final training objective of our ASSPH, where $\mu_{1}$ and $\mu_{2}$ are the trade-off parameters to balance different optimization goals.
\begin{equation}
\min\nolimits_{\theta^{I},\theta^{T}}\mathcal{L}=\mathcal{L}_{sr} + \mu _{1}\mathcal{L}_{cp} + \mu _{2}\mathcal{L}_{sa} \label{equ:L}
\end{equation}

\vspace{0.03in}
\noindent
\textbf{Asymmetric Similarity-Preserving Binary Optimization.} As our goal is to optimize the semantic loss $\mathcal{L}(B^{I},B^{T})$ between binary hash codes, we design the asymmetric binary optimization to constrain the hidden semantic space and the binary hashing space togethers. Traditional work mainly optimizes hash coding by reducing the quantization loss from real value to binary value. Still, gradient descent based algorithms in symmetric hashing learning will force the pair-wise instance to be updated towards the same directions and switch the sign of the hash code in the worst case~\cite{huang2019accelerate}. 

Inspired by the asymmetric hashing methods~\cite{zhang2019scalable,meng2020asymmetric}, we propose a two-stage asymmetric binary similarity optimization strategy and stabilize the hash learning by optimizing one side of the sub-network at each stage. Specifically, we optimize the continuous coding of one modality with the help of the binary coding of the other modality using $\mathcal{L}(H^{I},B^{T})$ and $\mathcal{L}(B^{I},H^{T})$, which also ensures the consistency of both the structure and the semantics of the hash space.

Besides, we also follow the previous approximation strategy~\cite{liu2020joint} to reduce the quantization loss by taking advantage of the convergence of the $tanh()$ function to $sgn()$. We replace the activation function of the final layer with $tanh()$ and adopt the following strategy $h_{i} = tanh(\eta x_{i}), \eta \to +\infty $. During the training stage, $\eta $ will gradually increase with the increase of epoch.

\begin{algorithm}[tb]
	\LinesNumbered
	\KwIn{Training set $\{I_{k}, T_{k}\}_{k=1}^{M}$and their corresponding features $F^{I}$ and $F^{T}$;  batch size $m$; train epoch $E$, $NN_{K_R}$ scale $K_R$ and $NN_{K_S}$ scale $K_S$.}
	\KwOut{Hashing function $ImgNet(F^{I},\theta^{I})$ and $TxtNet(F^{T},\theta^{T})$}
	Initialize $\theta^{I}$, $\theta^{T}$, epoch $e \gets 0$, $S\gets$ Eq.~\eqref{S_CAL}, $R\gets$ Eq.~\eqref{R_CAL};\\
	\For{each $e \in [1, E]$}
	{
		\For {$\left \lfloor \frac{M}{m} \right \rfloor iterations$}
	    {
	        Sample $m$ instances $\{f_{k}^{I}, f_{k}^{T}\}_{k=1}^{m}$ from training set;\\ 
	        Calculate the representations $H_{k}^{I}\gets ImgNet(f_{k}^{I},\theta^{I})$, $H_{k}^{T}\gets TxtNet(F_{k}^{T},\theta_{T})$ and get the corresponding $S(H^{I}, H^{T})$;\\
	        Update $\theta^{I}$ and $\theta^{T}$ using $\mathcal{L}(H^{I},H^{T})$;\\
	        Calculate $B^{I}\gets sgn(H^{I})$ and $B^{T}\gets sgn(H^{T})$;\\
	        Update $\theta^{I}$/$\theta^{T}$ using $\mathcal{L}(H^{I},B^{T})$/$\mathcal{L}(B^{I},H^{T})$;\\
	    }
		Update $R_{e}$ with Eq.~\eqref{R_UPDATE};\\
	}
	Return $ImgNet(F^{I},\theta^{I})$ and $TxtNet(F^{T},\theta^{T})$;\\
	\caption{Adaptive Structural Sim. Preserving Hashing}
	\label{alg:TEmax}
\end{algorithm}

\begin{figure*}[ht]
\centering  
\subfigure[I2T on NUS-WIDE]{
\label{Fig.pr.1}
\includegraphics[width=0.24\textwidth]{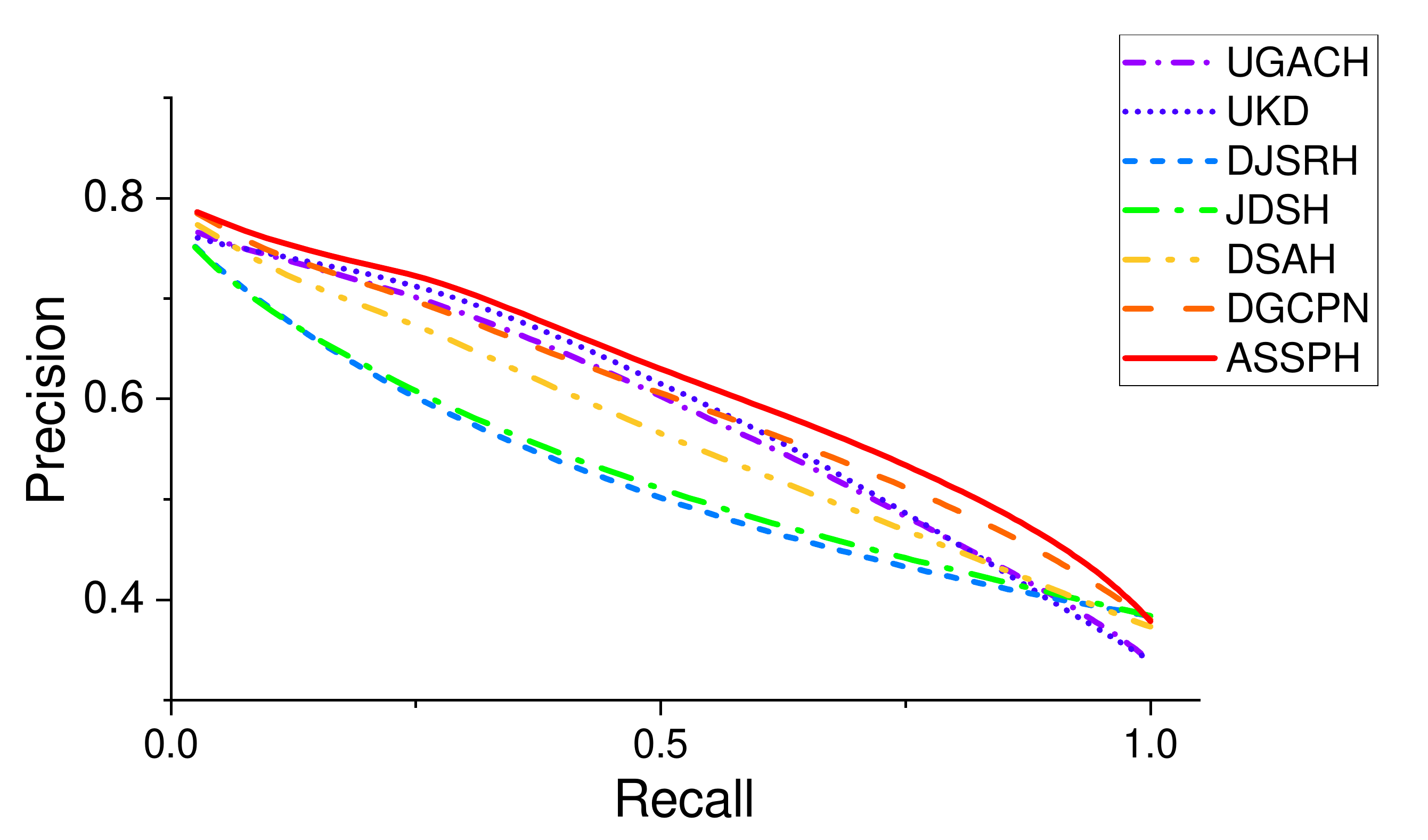}}
\subfigure[T2I on NUS-WIDE]{
\label{Fig.pr.2}
\includegraphics[width=0.24\textwidth]{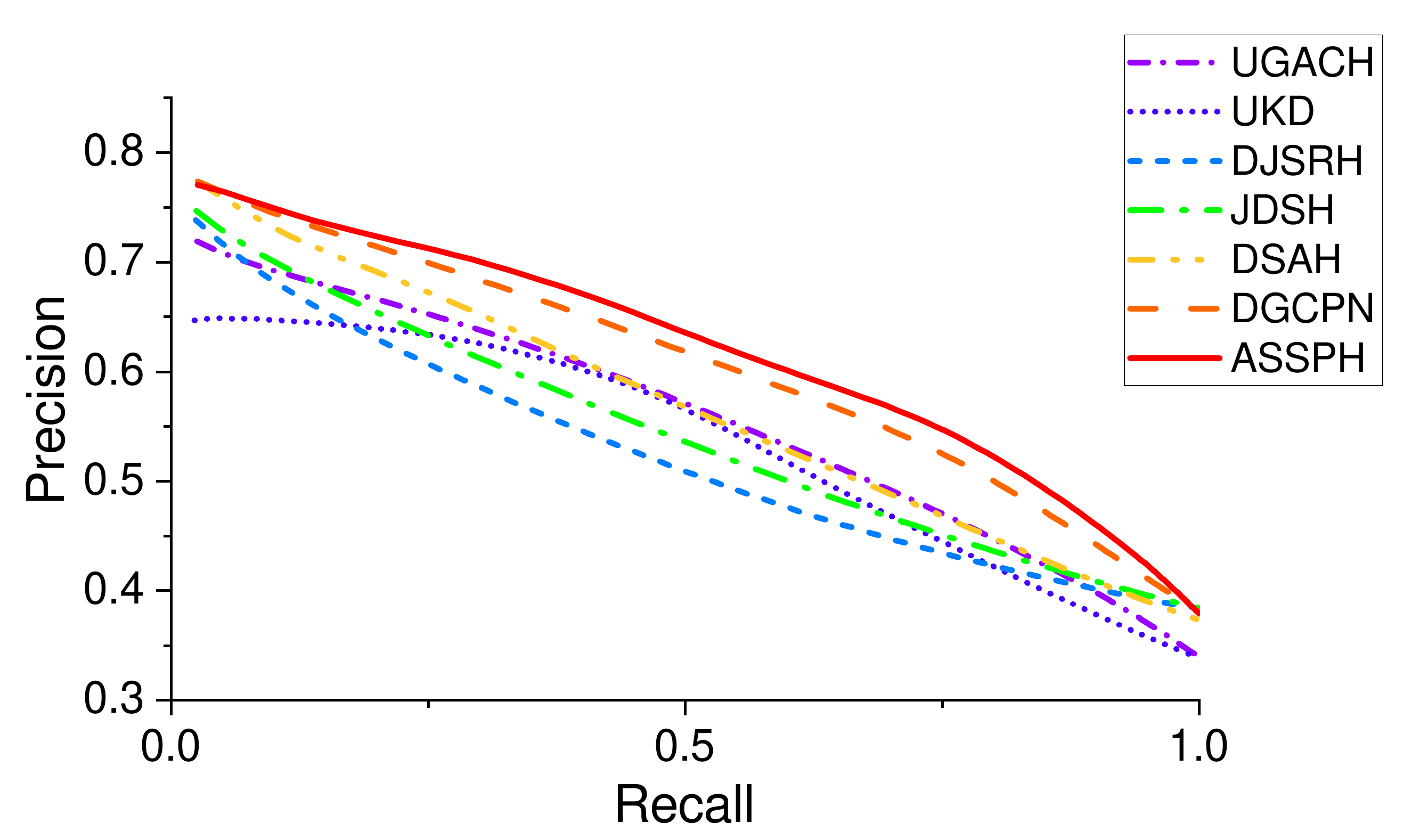}}
\subfigure[I2T on MIRFlickr-25K]{
\label{Fig.pr.3}
\includegraphics[width=0.24\textwidth]{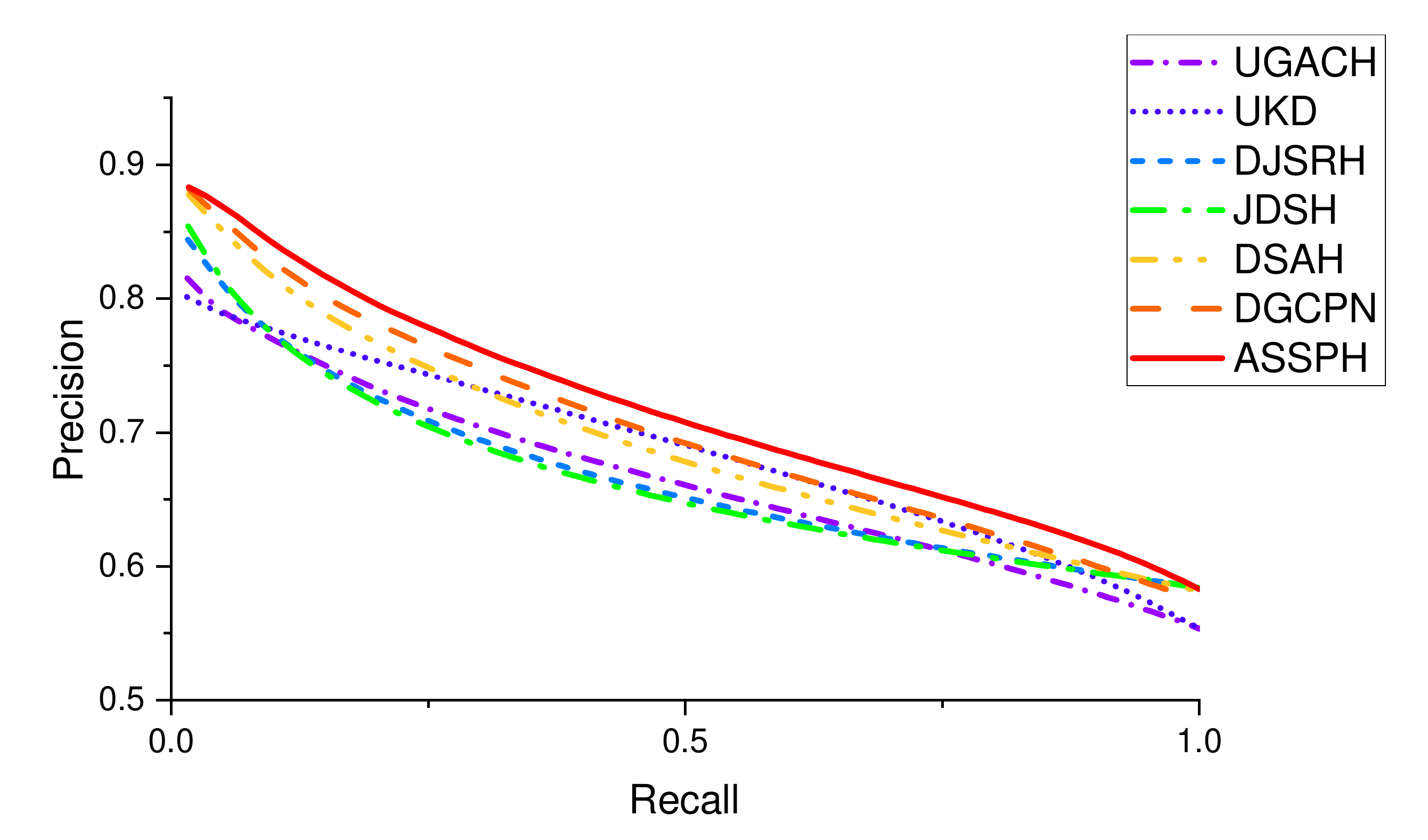}}
\subfigure[T2I on MIRFlickr-25K]{
\label{Fig.pr.4}
\includegraphics[width=0.24\textwidth]{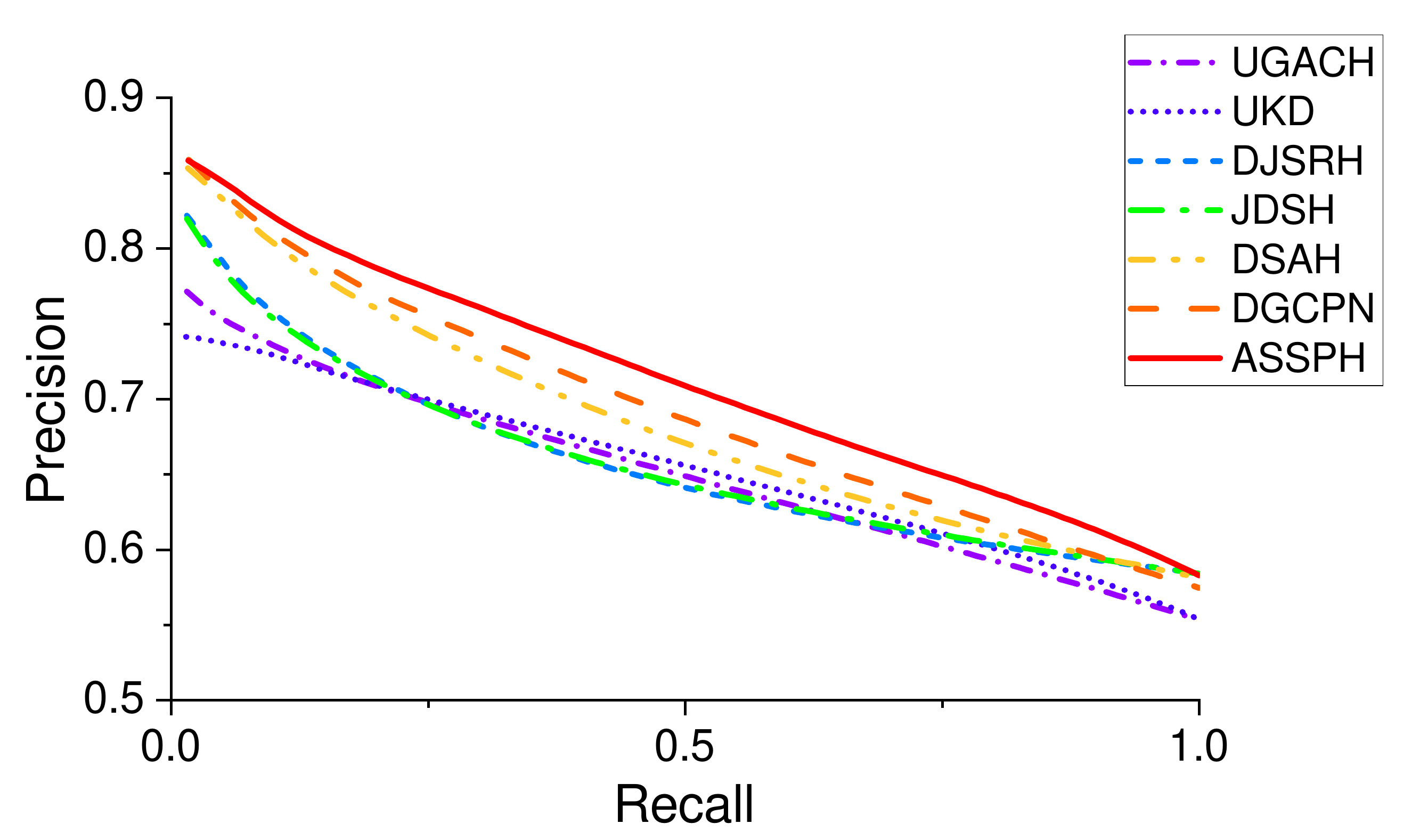}} \\
\vspace{-0.15in}
\subfigure[I2T on NUS-WIDE]{
\label{Fig.topk.1}
\includegraphics[width=0.24\textwidth]{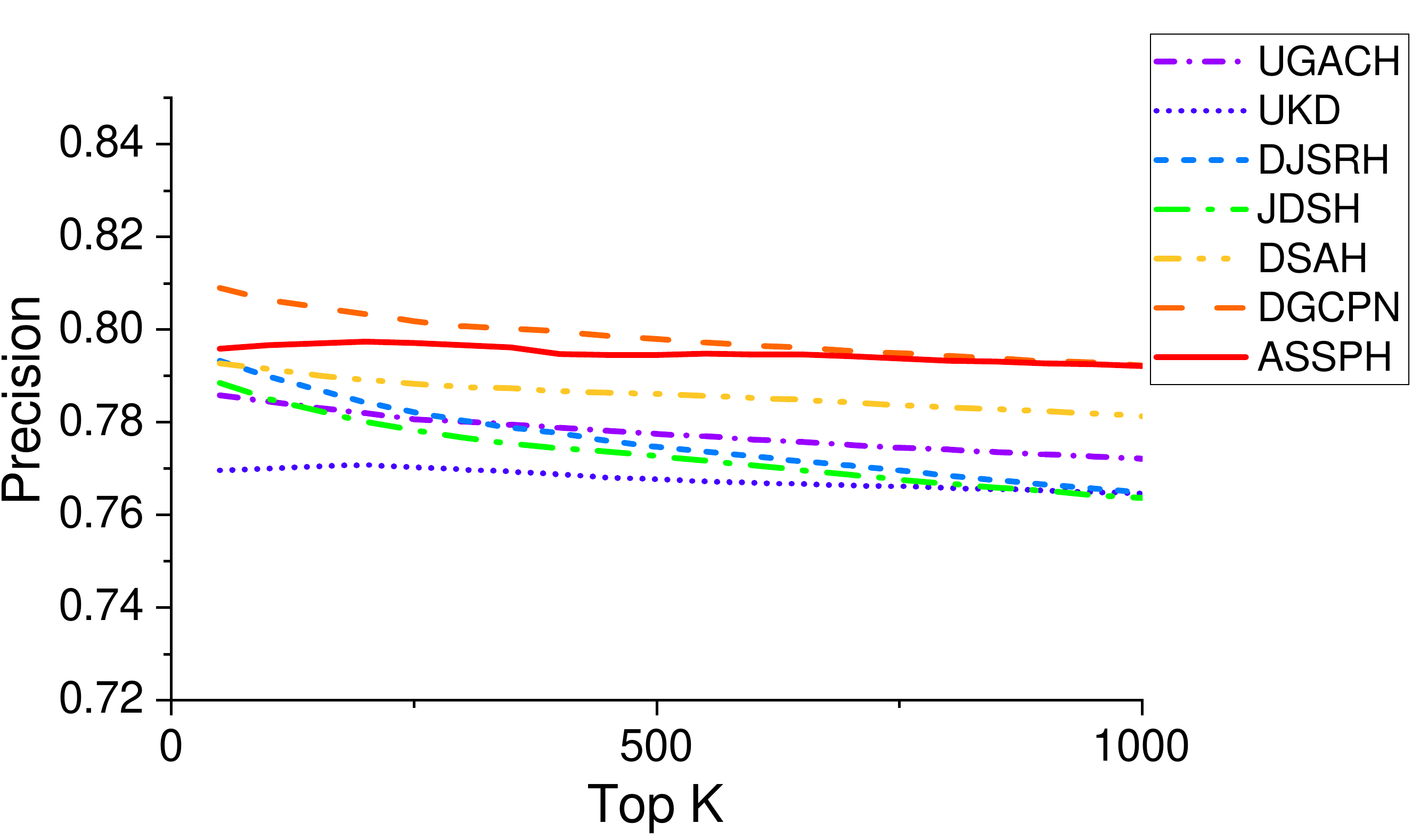}}
\subfigure[T2I on NUS-WIDE]{
\label{Fig.topk.2}
\includegraphics[width=0.24\textwidth]{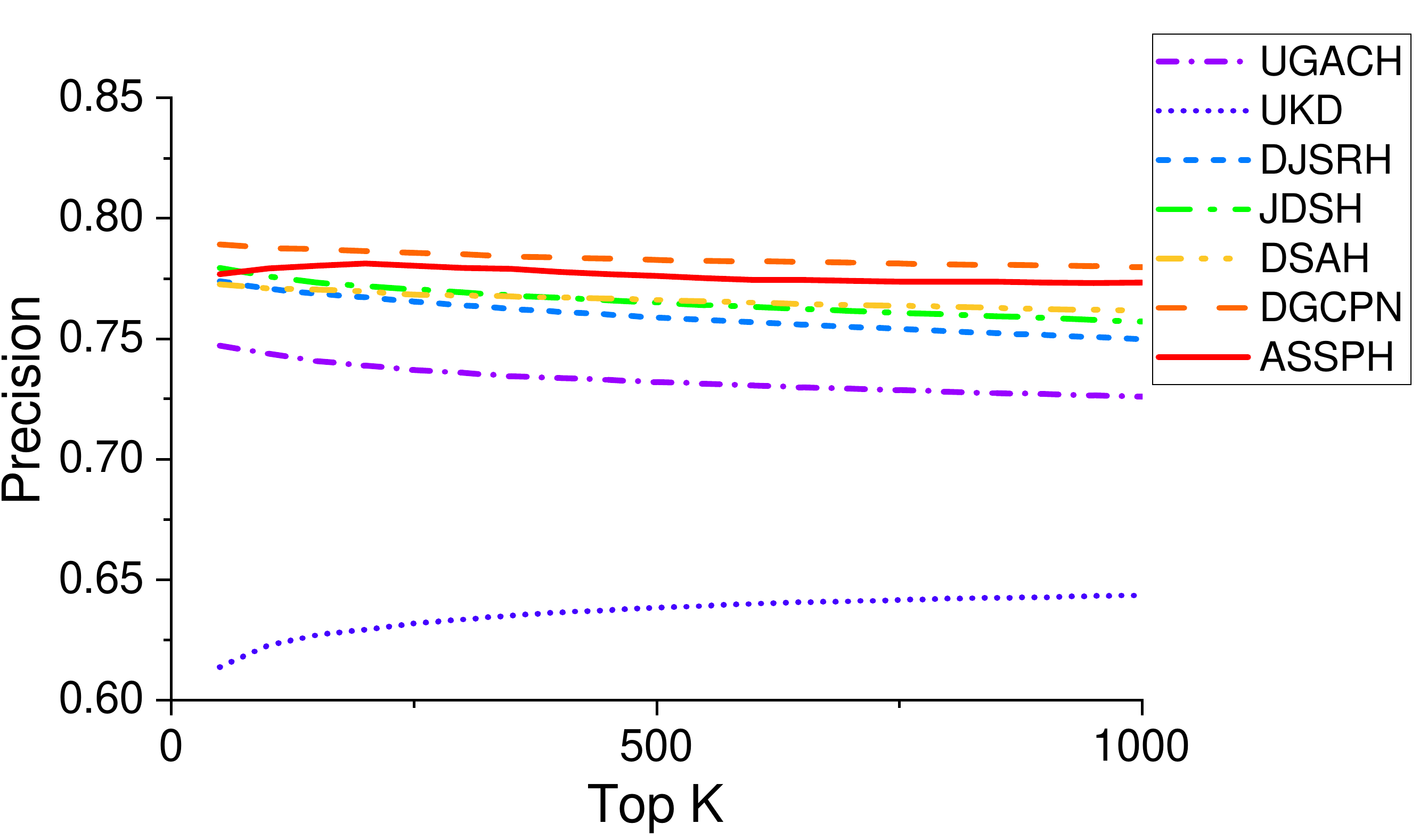}}
\subfigure[I2T on MIRFlickr-25K]{
\label{Fig.topk.3}
\includegraphics[width=0.24\textwidth]{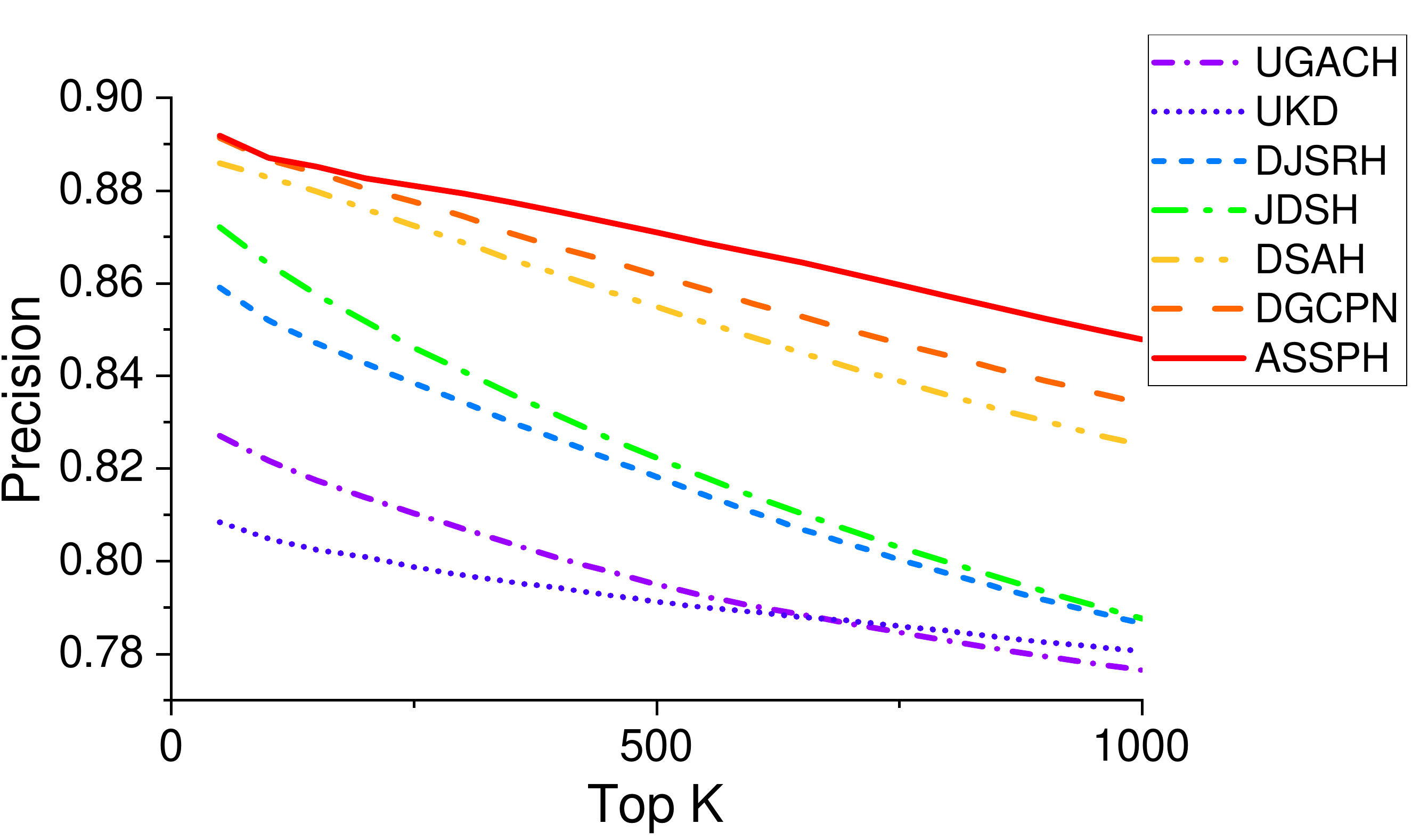}}
\subfigure[T2I on MIRFlickr-25K]{
\label{Fig.topk.4}
\includegraphics[width=0.24\textwidth]{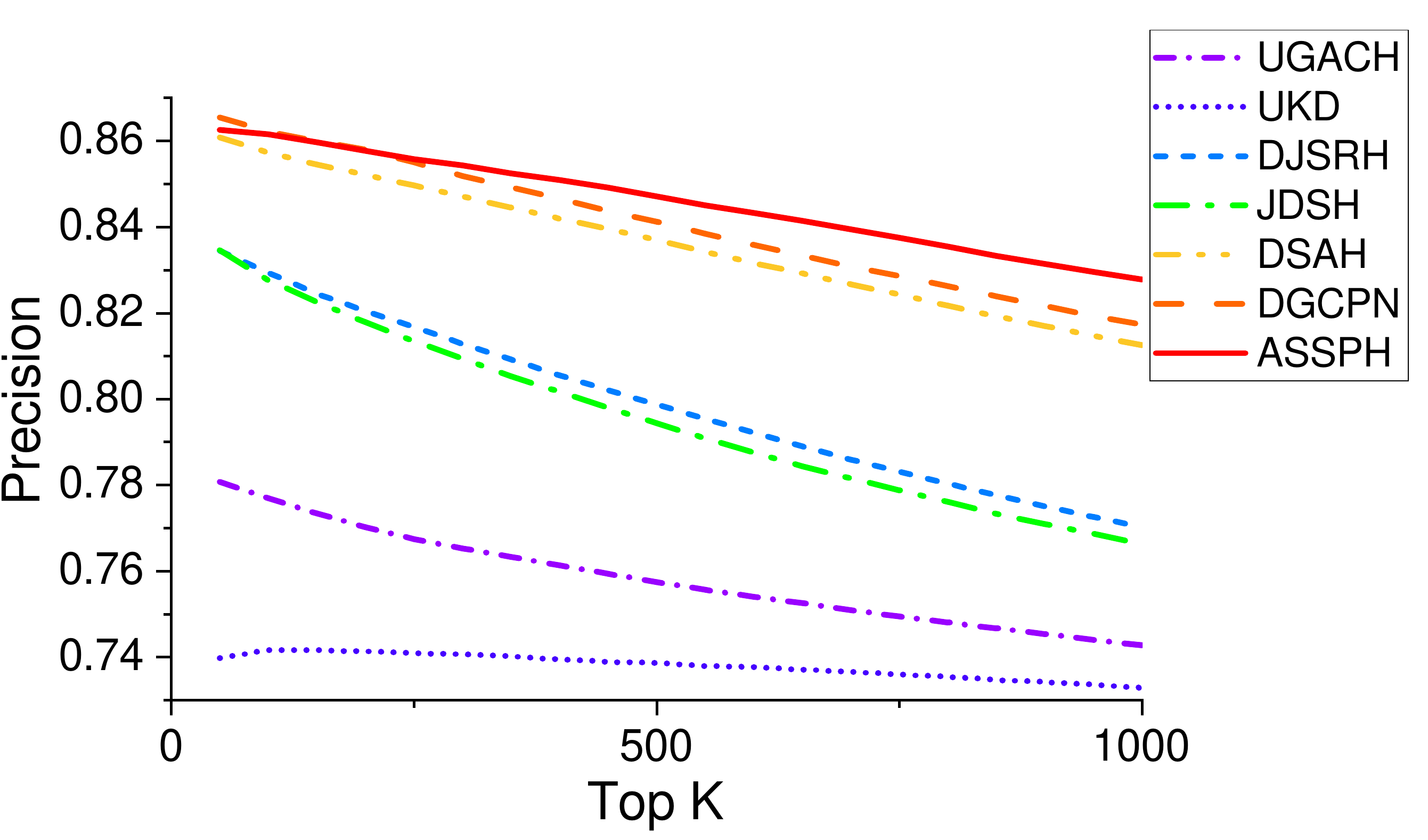}}
\vspace{-0.1in}
\caption{The precision-recall and Top-$k$ precision curves on NUS-WIDE and MIRFlickr-25K datasets with 64 bit hash code. 
}
\vspace{-0.1in}
\label{Fig.pr}
\end{figure*}

\section{Experimental Evaluation}

\subsection{Experiment Setup}

\vspace{0.03in} 
\noindent
\textbf{Datasets.} Two commonly used multi-label image-text cross-modal datasets, NUS-WIDE and MIRFlickr-25K, are chosen for our experiments. As a piece of unsupervised work, we do not use the label information in the dataset during the training process but only use it as a basis when verifying the performance. 
\emph{NUS-WIDE}~\cite{chua2009nus} consists of 269,648 labeled multimodal instances, each containing an image and a text description in pairs. Following previous work~\cite{wu2018unsupervised, su2019deep,liu2020joint}, we keep the instances corresponding to the top-10 most frequent labeled classes, resulting in 186,577 instances. 
%
\emph{MIRFlickr-25K}~\cite{huiskes2008mir} consists of 25,000 tagged multimodal instances collected from the Flickr website. Each instance contains an image and a corresponding text description, and they are divided into 24 categories in total. After removing the unlabeled instances, there are 20,015 instances left. 
%
In both datasets, we follow previous work~\cite{wu2018unsupervised,su2019deep,liu2020joint} to select 2,000 random instances as the query set and use the remaining instances to form the database, which also includes a training set consisting of 5,000 instances.
As for the single-label data set, it is a complex scene for our method because the positive correlation between data is much sparser than that of the multi-label scene, resulting in more errors in the unsupervised construction of the correlation. However, our model still achieves some improvements over existing methods. We have evaluated our ASSPH on the single-label Wiki dataset and please refer to the appendix for the details.


\vspace{0.03in} 
\noindent
\textbf{Evaluation Protocol.}
We verify the accuracy of hash retrieval in the task of mutual image and text retrieval, i.e., using an image as a query and retrieving the text associated with it (I2T) and vice versa (T2I).
To evaluate the retrieval quality, we adopt three standard evaluation metrics, including Mean Average Precision (MAP), Precision-Recall Curve (PR-Curve), and Top-$k$-Precision. 
At the same time, considering the actual application scenario, we also evaluate our ASSPH under MAP@50 and again please refer to the appendix for the detailed MAP@50 results. 


\vspace{0.03in} 
\noindent
\textbf{Baselines.} We implement eight state-of-the-art methods based on deep hash learning proposed in the last three years as competitors, including UGACH~\cite{zhang2018unsupervised}, UKD~\cite{hu2020creating}, SRCH~\cite{wang2020set}, DJSRH~\cite{su2019deep}, DSAH~\cite{yang2020deep}, JDSH~\cite{liu2020joint}, DGCPN~\cite{yu2021deep} and UCH~\cite{li2019coupled}.

\vspace{0.03in} 
\noindent
\textbf{Implementation Details.}
In our implementation, we take 4,096 dimensional features extracted from AlexNet for images and use the 1,000-dimensional BoW features for texts. For the hashing representation network, we directly use two nonlinear layers with 4,096 dimensions in the hidden layer.
For the competitors, we adopt their original implementations released by the authors and follow their best settings to perform the experiments. We also choose the same image and text embeddings (AlexNet and BoW features) for fair comparisons. The training epoch is uniformly set to 50 on both datasets, except UGACH and UKD with their own settings. Our method and competitors are implemented based on python 3.6 and trained on a single 2080ti GPU. 
Last but not least, we fix the batch size to 32 and use the SGD optimizer with 0.9 momentum and 0.0005 weight decay. The learning rates are set to 0.001. For other parameters in our ASSPH, we perform cross-validation and finally select the following set of parameters with the selection range in brackets: $K_{R}=50,[5,500]$, $K_{S}=2000,[1000,4000]$, $\mu_{1}=2[0.5,5]$, $\mu_{2}=1[0.5,5]$, $\beta=1.5,[1,3]$ and $\gamma=0.3[0,1]$. We use the same parameters on both datasets.

\begin{table*}[!ht]
\caption{The MAP performances of MIRFlickr-25K and NUS-WIDE at various hashing code lengths}
\label{MAP}
\vspace{-0.1in}
\centering
\begin{tabular}{|c||c|c|cccc|cccc|}
\hline
&     &      & \multicolumn{4}{c|}{MIRFlickr-25K}                                      & \multicolumn{4}{c|}{NUS-WIDE}               \\ \cline{4-11} 
& \multirow{-2}{*}{Task} & \multirow{-2}{*}{Method} & 16-bit                        & 32-bit                        & 64-bit                        & 128-bit                       & 16-bit                        & 32-bit                        & 64-bit                        & 128-bit                       \\ \hline
 \multirow{18}{*}{\rotatebox[origin=c]{90}{CNN backbone:AlexNet}}&            \multirow{9}{*}  {I2T}        & UGACH                    & 0.6595                                 & 0.6674                                 & 0.6777                                 & 0.6805                                 & 0.6107                                 & 0.6180                                 & 0.6167                                 & 0.6227                                 \\
&                       & UKD-SS                   & 0.6922                                 & 0.6903                                 & 0.6982                                 & 0.6963                                 & 0.6026                                 & 0.6202                                 & 0.6264                                 & 0.6112                                 \\
&                       & $UCH^{*}$                      & 0.6540                                  & 0.6690                                  & 0.6790                                  & -                                      & -                                      & -                                      & -                                      & -                                      \\
&                       & $SRCH^{*}$                     & 0.6808                                 & 0.6916                                 & 0.6997                                 & -                                      & 0.5441                                 & 0.5565                                 & 0.5671                                 & -                                      \\
&                       & DJSRH                    & 0.6576                                 & 0.6606                                 & 0.6758                                 & 0.6856                                 & 0.5073                                 & 0.5239                                 & 0.5422                                 & 0.5530                                 \\
&                       & JDSH                     & 0.6417                                 & 0.6567                                 & 0.6731                                 & 0.6850                                 & 0.5128                                 & 0.5170                                 & 0.5475                                 & 0.5563                                 \\
&                       & DSAH                     & 0.6914                                 & 0.6981                                 & 0.7035                                 & 0.7086                                 & 0.5677                                 & 0.5839                                 & 0.5959                                 & 0.6053                                 \\
&                       & DGCPN                    & 0.6991                                 & 0.7116                                 & 0.715                                  & 0.7243                                 & 0.5993                                 & 0.6150                                 & 0.6262                                 & 0.6338                                 \\
 &  & \textbf{ASSPH}           & {
\color[HTML]{FE0000} \underline{0.7138}} & {\color[HTML]{FE0000} \underline{0.7244}} & {\color[HTML]{FE0000} \underline{0.7301}} & {\color[HTML]{FE0000} \underline{0.7341}} & {\color[HTML]{FE0000} \underline{0.6238}} & {\color[HTML]{FE0000} \underline{0.6340}} & {\color[HTML]{FE0000} \underline{0.6477}} & {\color[HTML]{FE0000} \underline{0.6512}} \\ \cline{2-10}
&  \multirow{9}{*}{T2I}                     & UGACH                    & 0.6512                                 & 0.6587                                 & 0.6677                                 & 0.6693                                 & 0.5993                                 & 0.6022                                 & 0.6019                                 & 0.6006                                 \\
&                       & UKD-SS                   & 0.6760                                 & 0.6799                                 & 0.6689                                 & 0.6722                                 & 0.5830                                 & 0.6042                                 & 0.5963                                 & 0.6075                                 \\
&                       & $UCH^{*}$                      & 0.6610                                  & 0.6670                                  & 0.6680                                  & -                                      & -                                      & -                                      & -                                      & -                                      \\
&                       & $SRCH^{*}$                     & 0.6971                                 & 0.7081                                 & 0.7146                                 & -                                      & 0.5533                                 & 0.5670                                  & 0.5754                                 & -                                      \\
&                       & DJSRH                    & 0.6594                                 & 0.6528                                 & 0.6688                                 & 0.6801                                 & 0.4936                                 & 0.5325                                 & 0.5508                                 & 0.5470                                 \\
&                       & JDSH                     & 0.6352                                 & 0.6608                                 & 0.6686                                 & 0.6736                                 & 0.5153                                 & 0.4982                                 & 0.5683                                 & 0.5704                                 \\
 &                      & DSAH                     & 0.6903                                 & 0.6916                                 & 0.7013                                 & 0.7038                                 & 0.5695                                 & 0.5874                                 & 0.6066                                 & 0.6114                                 \\
 &                      & DGCPN                    & 0.6991                                 & 0.7074                                 & 0.7133                                 & 0.7233                                 & 0.6141                                 & 0.6353                                 & 0.6456                                 & 0.6522                                 \\
& & \textbf{ASSPH}           & {\color[HTML]{FE0000} \underline{0.7170}} & {\color[HTML]{FE0000} \underline{0.7265}} & {\color[HTML]{FE0000} \underline{0.7315}} & {\color[HTML]{FE0000} \underline{0.7380}} & {\color[HTML]{FE0000} \underline{0.6332}} & {\color[HTML]{FE0000} \underline{0.6502}} & {\color[HTML]{FE0000} \underline{0.6646}} & {\color[HTML]{FE0000} \underline{0.6628}} \\ \hline\hline
\multirow{12}{*}{\rotatebox[origin=c]{90}{CNN Backbone: VGG-16}} 
   &\multirow{6}{*}{I2T} & UGACH                    & 0.685                                 & 0.693                                 & 0.704                                 & 0.702                                 & 0.613                                 & 0.623                                 & 0.628                                 & 0.631                                 \\
 &                      & UKD-SS                   & 0.714                                 & 0.718                                 & 0.725                                 & 0.720                                 & 0.614                                 & 0.637                                 & 0.638                                 & 0.645                                 \\
 &                      & UCH                      & 0.654                                 & 0.669                                 & 0.679                                 & -                                     & -                                     & -                                     & -                                     & -                                     \\
 &                      & SRCH                     & 0.6808                                & 0.6916                                & 0.6997                                & -                                     & 0.5441                                & 0.5565                                & 0.5671                                & -                                     \\
 &                      & DGCPN                    & 0.732                                 & 0.742                                 & 0.751                                 & -                                     & 0.625                                 & 0.635                                 & 0.654                                 & -                                     \\
& & \textbf{ASSPH}           & {\color[HTML]{FE0000} \textbf{0.739}} & {\color[HTML]{FE0000} \textbf{0.753}} & {\color[HTML]{FE0000} \textbf{0.757}} & {\color[HTML]{FE0000} \textbf{0.762}} & {\color[HTML]{FE0000} \textbf{0.639}} & {\color[HTML]{FE0000} \textbf{0.660}} & {\color[HTML]{FE0000} \textbf{0.667}} & {\color[HTML]{FE0000} \textbf{0.669}} \\ \cline{2-11}
 &\multirow{6}{*}{T2I}                      & UGACH                    & 0.673                                 & 0.676                                 & 0.686                                 & 0.690                                 & 0.603                                 & 0.614                                 & 0.640                                 & 0.641                                 \\
  &                     & UKD-SS                   & 0.715                                 & 0.716                                 & 0.721                                 & 0.719                                 & 0.630                                 & 0.656                                 & 0.657                                 & 0.663                                 \\
  &                     & UCH                      & 0.661                                     & 0.667                                     & 0.668                                     & -                                     & -                                     & -                                     & -                                     & -                                     \\
    &                   & SRCH                     & 0.6971                                & 0.7081                                & 0.7146                                & -                                     & 0.5533                                & 0.567                                 & 0.5754                                & -                                     \\
   &                    & DGCPN                    & 0.729                                 & 0.741                                 & 0.749                                 & -                                     & 0.631                                 & 0.648                                 & 0.660                                 & -                                     \\
& & \textbf{ASSPH}           & {\color[HTML]{FE0000} \textbf{0.746}} & {\color[HTML]{FE0000} \textbf{0.756}} & {\color[HTML]{FE0000} \textbf{0.764}} & {\color[HTML]{FE0000} \textbf{0.767}} & {\color[HTML]{FE0000} \textbf{0.652}} & {\color[HTML]{FE0000} \textbf{0.673}} & {\color[HTML]{FE0000} \textbf{0.676}} & {\color[HTML]{FE0000} \textbf{0.674}} \\ \hline
\end{tabular}%
\vspace{-0.15in}
\end{table*}

\vspace{-0.03in}
\subsection{Retrieval Performance}

We compare the performance of ASSPH againt its competitors. Note that since the implementations of UCH and SRCH have not been released, we directly copy the MAP results from the original papers without PR-Curve and Top-$k$ Precision Curve. 

First, we report in the top part of Table \ref{MAP} the MAP performance (averaged over three times) based on different code lengths for two tasks, I2T and T2I. Our ASSPH consistently performs the best. 

Then, we report PR-Curve (without the two data points when recall is 0 and recall is 1) and Top-$k$ in Figure~\ref{Fig.pr} with 64-bit code length. As expected, our method also outperforms all the competitors in the PR curves for both tasks. For the Top-$k$ indicator, when $k$ is small, our improvement is not obvious, but as $k$ increases, our ASSPH performs better. This is because these original static similarity metrics in competitors are reliable in top-ranked results. 
The semantic metric reconstructed in DJSRH, JDSH and DGCPN depends heavily on the accuracy of the original semantic metric, while the current deep learning features have a high confidence level in top-ranked results (e.g., Top-50). In other words, they can only take advantage of the relationship between a few data (such as the relationship between the top-50 related data), and most of the relationships can only be filtered out as noise. 
However, ASSPH can make full use of the relationship between all data. While the improvement of ASSPH may be relatively limited in the top-50 retrieval performance, ASSPH can instead optimize the overall retrieval results.

Next, we report some results with a different CNN backbone. Some unsupervised hash works~\cite{zhang2018unsupervised,yu2021deep} use VGG to get the image features with higher quality to improve the effectiveness of unsupervised works. Accordingly, we report the retrieval performance of different methods based on VGG-16 in the bottom part of Table~\ref{MAP}.
For the convenience of comparison, we directly list the experimental results reported by each competitor in the original literature. Note that DJSRH, JDSH, and DSAH are missing here because their original papers only report the results of MAP@50, and the metric we compare are based on MAP@all. Using AlexNet will result in a 2\% to 4\% reduction in the overall effect of the experiments compared with the performance in the top part of Table~\ref{MAP}. This is the main reason why our results reported in the top part of Table~\ref{MAP} 
deviate from the results in their original literature. 

We observe that in the experiment based on VGG, our improvement is smaller than that based on AlexNet. This is because the prior knowledge used as the training targets in existing work, such as the neighboring network in UGACH~\cite{zhang2018unsupervised} or the cross-mode similarity measure in DGCPN~\cite{yu2021deep}, will be more reliable when the feature quality improves. However, our method still outperforms its competitors regardless of the original feature quality. The difference between the performance with different backbone models can further reflect the versatility of ASSPH, i.e., it remains effective even when the original similarity metric has low quality.


\vspace{-0.03in}
\subsection{Ablation Studies}

We design several variants of ASSPH to show the effectiveness of several key modules.
To be more specific, we implement four variants models, including 
i) \textbf{ASSPH\_NoAdapt} that removes the adaptive learning strategy from ASSPH, so $R$ will remain unchanged during the whole training process; ii) \textbf{ASSPH\_PairCorr} that replaces the structural correlation with the pairwise correlation; iii) \textbf{ASSPH\_NoCorr} that removes the correlational relationship and the corresponding correlation preserving loss $\mathcal{L}_{cp}$;
and iv) \textbf{ASSPH\_NoBinOpt} that removes the similarity-preserving binary optimization strategy from ASSPH.

The performance comparison among different variants is reported in Table~\ref{tab:freq1}. In most cases, we can observe that these four key modules effectively improve the performance by different extents and the results fully illustrate their effectiveness in ASSPH. For the BinOpt module, it is introduced to speedup the convergence process to the optimal binary representation. As the training on MIRFlickr tends to converge quickly, our BinOpt module does not improve much in this fast-converging scenario, so ASSPH performs  only slightly better than ASSPH\_NoBinOpt.

\vspace{-0.05in}
\subsection{Parameter analysis}

In this set of experiments, we conduct further analysis on the effects of different parameters in our method, including the $K_{S}$ and $\gamma$ in structural similarity construction, $K_{R}$ in correlational relationship mining and the training batch size. Parameter $\beta$, a balance parameter to adjust the data bias in Eq.~(\ref{LOSS3}), has a rather small optional range (e.g., [1,2]). We follow 
previous works~\cite{su2019deep,liu2020joint,yang2020deep,yu2021deep} to set $\beta$ to 1.5, and perform cross-validation to confirm its plausibility.

\begin{figure}[h]
\vspace{-0.1in}
\centering  
\subfigure[MAP on NUS-WIDE]{
\label{Fig.Nparam.1}
\includegraphics[width=0.22\textwidth]{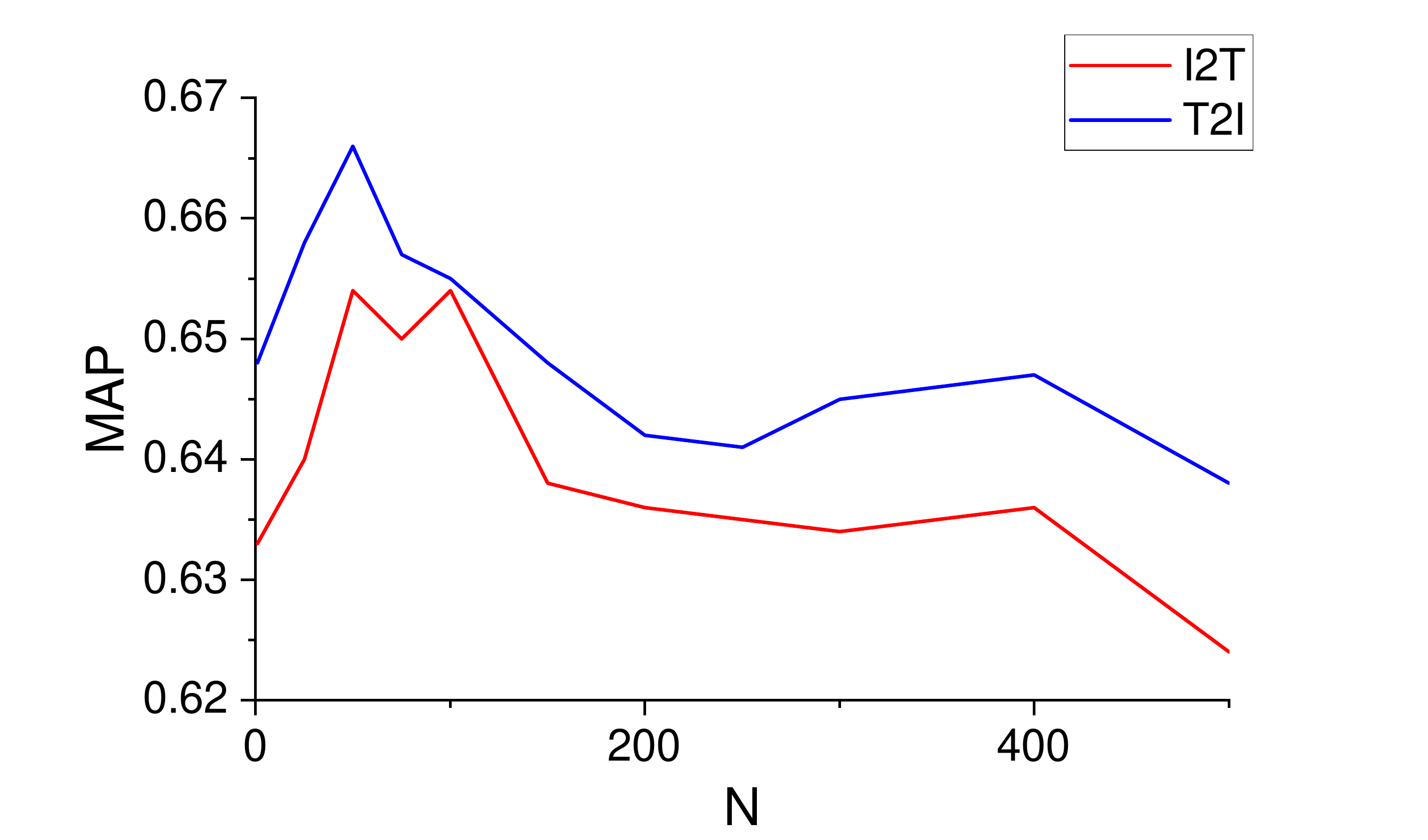}}
\subfigure[MAP on MIRFlickr-25K]{
\label{Fig.Nparam.2}
\includegraphics[width=0.22\textwidth]{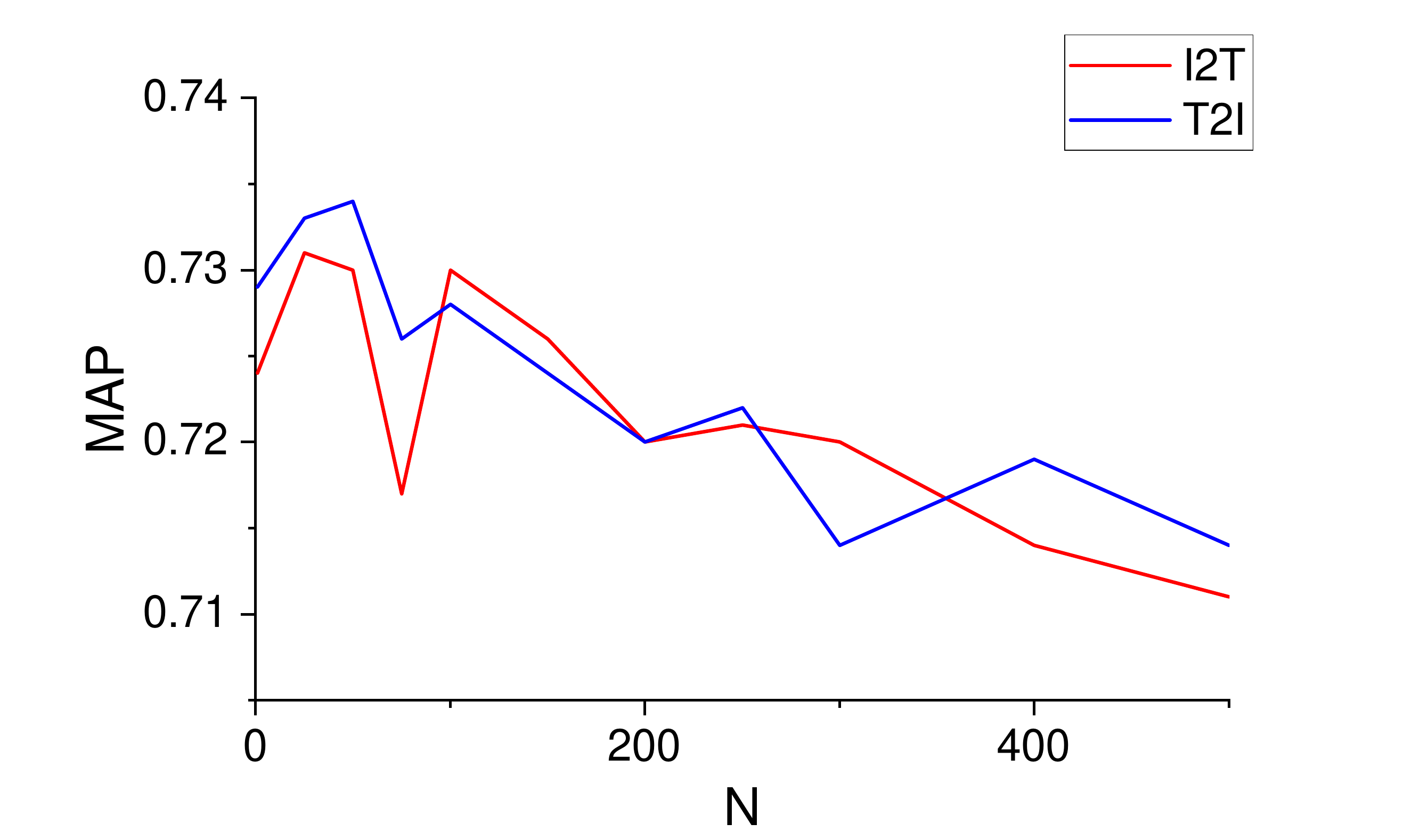}}
\vspace{-0.2in}
\caption{The MAP performance with different $K_{R}=N$ on NUS-WIDE and MIRFlickr-25K datasets (64 bit hash code).}
\label{Fig.Nparam}
\vspace{-0.2in}
\end{figure}

First, we study the value of $K_{R}$, denoted by $N$, that records the neighborhood range used as positive samples in correlation mining. Figure~\ref{Fig.Nparam} shows the MAP performance achieved by our method under different $N$ values. As it is a single-peaked function of neighbors, we can see that when the adaptive learning strategy is effective under proper settings and our ASSPH can increase positive relations further while maintaining its reliability.

Next, we study the value of $K_{S}$ in Table~\ref{tab:MAP_parameters}. The table shows the cross-modal retrieval performance under different values of $K_S$. We can observe that, on the one hand, the final result is not sensitive to the value of $K_S$, and on the other hand, when $K_S$ increases significantly, the final result decreases. This is because as $K_S$ is approaching the training set size, we actually consider the correlation between all instances, which undoubtedly introduces additional noise because the similarity relations between distant instances are less credible.

In the same table, we also report the results of our experiments with different batch sizes/$\gamma$ values. As ASSPH has a low dependence on batch size, it allows the unsupervised learning strategy to learn effectively even in small samples and limited datasets. As the batch size increases from its default value 32 to 256, the performance changes are small. Parameter $\gamma$ controls the weight of structural similarity when calculating the structural semantic similarity $S$. We can observe that our method is still effective even when $\gamma=0$ (i.e., $S$ is purely based on structural similarity). However, combining structural similarity and static cross-modal similarity as a multi-level cross-mode similarity can further improve the performance.
%

\vspace{-0.03in}
\subsection{Further Analysis of the Adaptive Process}



\begin{figure}[h]
\vspace{-0.15in}
\centering  
\subfigure[Number of correlations during the training]{
\label{Fig.Adap.1}
\includegraphics[width=0.22\textwidth]{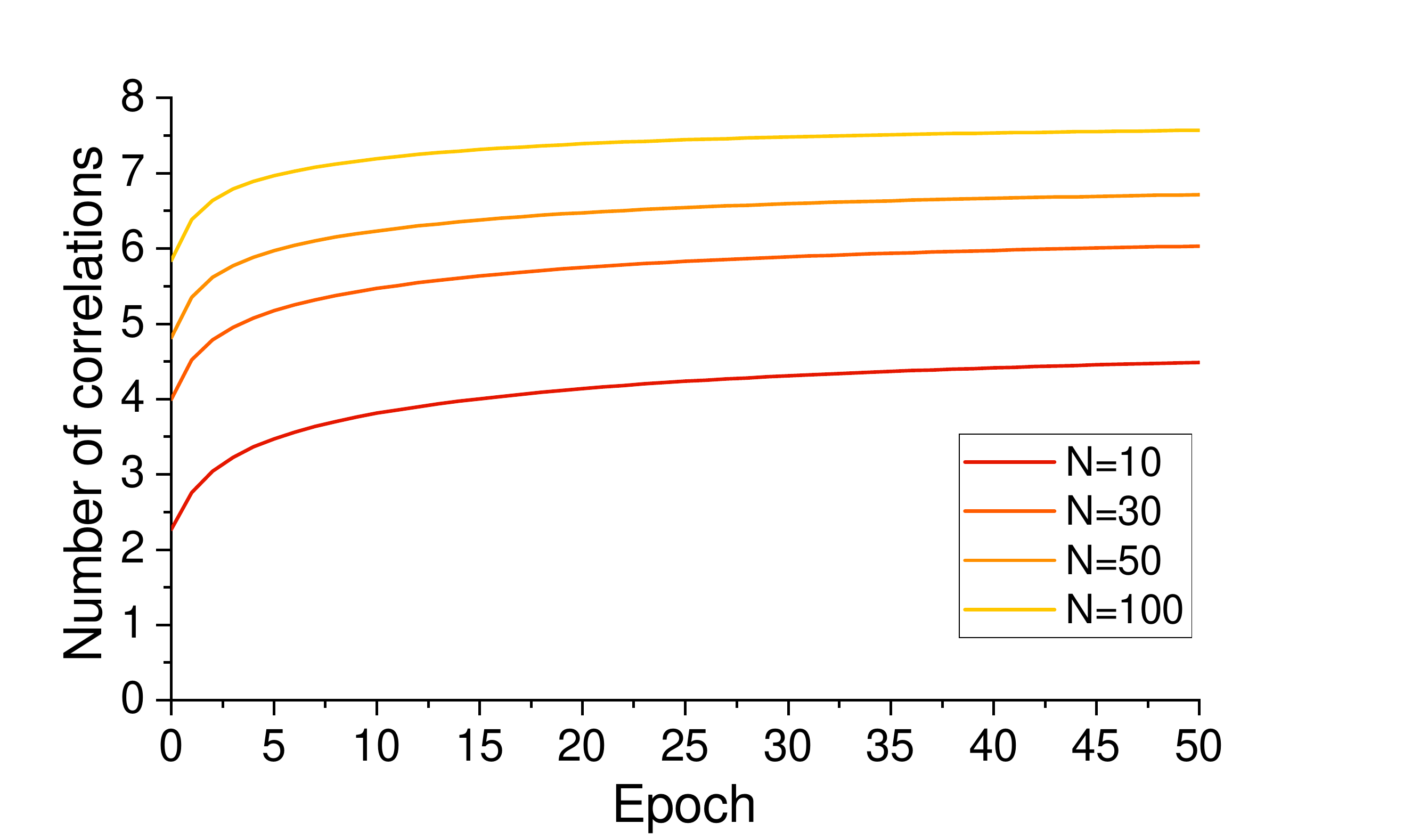}}
\subfigure[Distribution of Similarity between correlated instances]{
\label{Fig.Adap.2}
\includegraphics[width=0.22\textwidth]{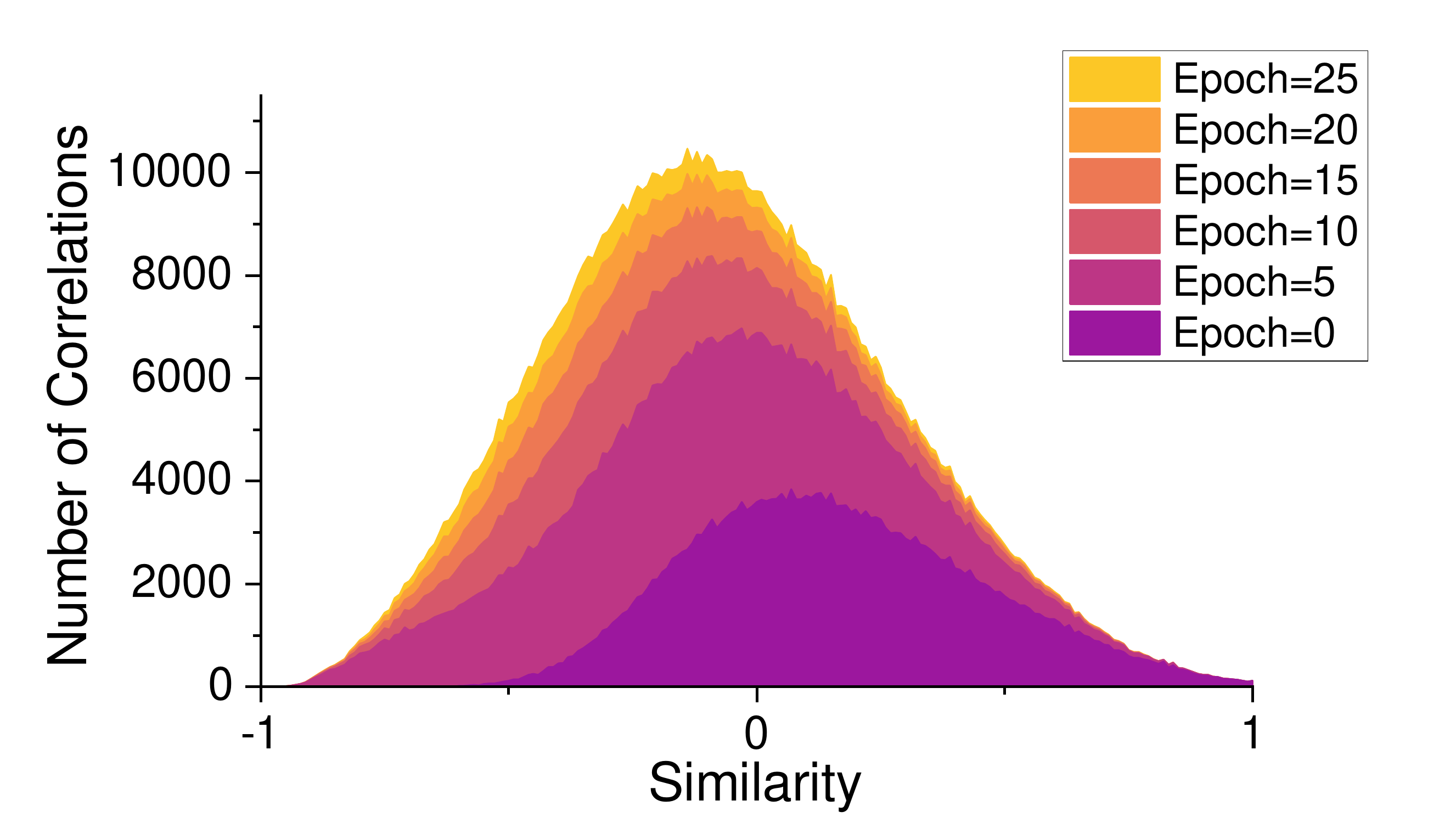}}
\vspace{-0.2in}
\caption{Descriptions of the adaptive correlation mining on MIRFlickr-25K with 64 bit hash code.}
\label{Fig.Adap}
\vspace{-0.2in}
\end{figure}

Figure~\ref{Fig.Adap} respectively describes the convergence of the adaptive mining process and the cross-modal semantic complement brought by it. Figure~\ref{Fig.Adap.1} shows the variation of the number of correlated instances under different scales $K_{R}=N$, where $Num$ on y-axis indicates that there are $2^{Num}\times 10^{4}$ instances. We can observe that our adaptive mining strategy expands the correlation set 4 times larger than the initial one. At the same time, the growth tends to be zero, which ensures the convergence of the optimization process. Figure~\ref{Fig.Adap.2} reports the variation of the similarity distribution among the correlated instances mined during the training process. The number of correlations increases with the training process while their similarity distribution gradually shifts to a minor direction. Meanwhile, we test the accuracy of the correlation relationships, and it rises from the initial 0.861 to 0.885 after 25 training epochs, which indicates the reliability of our adaptive learning.
Therefore, we can conclude that ASSPH obtains additional joint-modal correlations independent of the original static metrics, while most initial correlations are of higher similarity.  
These low-similarity correlations can be seen as hard samples that contribute to the enrichment of the semantic representation of our hashing model.

\begin{table}[h]
\vspace{-0.1in}
\caption{Ablations experiments 
with 64-bit hashcode}
\label{tab:freq1}
\vspace{-0.12in}
\begin{tabular}{ccccc}
\hline
\multirow{2}{*}{Method} & \multicolumn{2}{c}{NUS-WIDE} & \multicolumn{2}{c}{MIRFlickr-25K} \\ \cline{2-5} 
                        & I2T          & T2I          & I2T           & T2I           \\ \hline
ASSPH                   & 0.650         & 0.667        & 0.730          & 0.732         \\
ASSPH\_NoAdapt          & 0.638        & 0.654        & 0.723         & 0.729         \\
ASSPH\_PairCorr            & 0.633        & 0.654        & 0.718         & 0.728         \\
ASSPH\_NoCorr           & 0.604        & 0.617        & 0.696         & 0.691         \\
ASSPH\_NoBinOpt         & 0.639        & 0.648        & 0.725         & 0.732         \\ \hline
\end{tabular}
\vspace{-0.15in}
\end{table}
\begin{table}[h]
\caption{The MAP performance with different parameters}
\label{tab:MAP_parameters}
\vspace{-0.15in}
\begin{tabular}{c|c||cccc}
\hline
\multicolumn{2}{c}{} & \multicolumn{2}{c}{NUS-WIDE} & \multicolumn{2}{c}{MIRFlickr-25K} \\ \cline{3-6} 
\multicolumn{2}{c}{}    & I2T          & T2I          & I2T           & T2I           \\ \hline
\multirow{4}{*}{\rotatebox[origin=c]{90}{$K_S$}} & 
1000                   & 0.642         & 0.656        & 0.727          & 0.726         \\
&2000          & 0.648        & 0.665        & 0.730         & 0.721         \\
&3000           & 0.639        & 0.654        & 0.726         & 0.728         \\
&4000           & 0.627        & 0.655        & 0.722         & 0.719         \\
\hline
\hline
\multirow{4}{*}{\rotatebox[origin=c]{90}{batch size}} &                        
32                   & 0.634         & 0.652        & 0.726          & 0.732         \\
&64          & 0.636        & 0.651        & 0.727         & 0.723         \\
&128           & 0.639        & 0.654        & 0.726         & 0.725         \\
&256           & 0.638        & 0.655        & 0.727         & 0.729         \\
\hline\hline
\multirow{6}{*}{\rotatebox[origin=c]{90}{$\gamma$}} & 0                   & 0.633         & 0.649        & 0.722          & 0.721         \\
&0.2          & 0.632        & 0.651        & 0.73         & 0.729         \\
&0.4           & 0.630        & 0.644        & 0.727         & 0.727         \\
&0.6           & 0.628        & 0.646        & 0.719         & 0.722         \\
&0.8           & 0.622        & 0.641        & 0.705         & 0.71         \\
&1           & 0.623        & 0.639        & 0.668         & 0.662         \\
\hline
\end{tabular}
\vspace{-0.15in}
\end{table}

\section{Conclusion}

This paper proposes a novel unsupervised cross-modal hashing framework, named \emph{Adaptive Structural Similarity Preserving Hash (ASSPH)}, and conducts extensive experiments to verify its effectiveness. The framework is based on asymmetric structural semantic preserving with additional adaptive correlation expansion and constraints to learn the joint-semantic relationship adaptively. Combining the two parts is also a good solution for avoiding training collapse caused by unbalanced samples and small datasets, while mitigating the limitation from static metric reconstruction in traditional unsupervised cross-modal hashing.  

\section{Acknowledgements}

This research is supported in part by the National Natural Science Foundation of China under grant 62172107 and the National Key Research and Development Program of China under grant 2018YFB0505000.

\bibliographystyle{ACM-Reference-Format}
\bibliography{sample-base}

\appendix

\section{Performance on Wikipedia}
In this part, we test the MAP performance of our ASSPH on the Wikipedia dataset and show the comparison with some contrasting methods.

\subsection{Dataset}
The Wikipedia is a single-label dataset that consists of $2,866$ image-text pairs from 10 categories. We split the whole dataset into the retrieval/query set with 2173/693 image-text pairs, and use the whole retrieval set for training.

\subsection{Implementation Details}

Same as the comparison work, we adopt VGG-16 as the image feature extraction network and BoW as the text feature.
We fix the batch size as 32 and use the SGD optimizer with 0.9 momentum and 0.0005 weight decay. The learning rate is set to 0.001. For other parameters in our ASSPH, we perform cross-validation and finally select the following set of parameters with the selection range in brackets: $K_{R}=10,[5,100]$, $K_{S}=400,[100,1000]$, $\mu_{1}=1[0.5,5]$, $\mu_{2}=1[0.5,5]$ and $\gamma=0.2[0,1]$.

\subsection{Retrieval Performance}

We report the MAP performance on the Wikipedia dataset in Table~\ref{TAB_1}. Although our model ASSPH improves the average performance, the improvement is limited. This is beacuse single-label data set is a complex scene for our method as the positive correlation between data is much sparser than that of the multi-label scene, resulting in more errors in the unsupervised construction of the correlation. Especially when the hash code has only 16 bits, the first few rounds of adaptive optimization may introduce lots of noise, 
which leads to a poor performance. For the task of I2T, our model consistently outperforms the existing two baselines that report MAP@all in their original papers. For the task of T2I, though our model is not the best performer, it achieves a comparable performance as the best performer. 



\begin{table}[h]
\caption{The MAP performances of Wikipedia at various hashing code lengths with VGG-16}
\vspace{-0.1in}
\label{TAB_1}
\begin{tabular}{|c|c|ccc|}
\hline
\multirow{2}{*}{Task} & \multirow{2}{*}{Method} & \multicolumn{3}{c|}{Wikipedia}                                          \\ \cline{3-5} 
                      &                         & \multicolumn{1}{c|}{16-bit} & \multicolumn{1}{c|}{32-bit} & 64-bit \\ \hline
\multirow{3}{*}{I2T}  & SRCH                    & \multicolumn{1}{c|}{0.374}  & \multicolumn{1}{c|}{0.38}   & 0.391  \\ \cline{2-5} 
                      & DGCPN                   & \multicolumn{1}{c|}{0.404}  & \multicolumn{1}{c|}{0.413}  & 0.420   \\ \cline{2-5} 
                      & ASSPH                   & \multicolumn{1}{c|}{0.415}  & \multicolumn{1}{c|}{0.429}  & 0.435  \\ \hline \hline
\multirow{3}{*}{T2I}  & SRCH                    & \multicolumn{1}{c|}{0.376}  & \multicolumn{1}{c|}{0.401}  & 0.406  \\ \cline{2-5} 
                      & DGCPN                   & \multicolumn{1}{c|}{0.539}  & \multicolumn{1}{c|}{0.550}   & 0.558  \\ \cline{2-5} 
                      & ASSPH                   & \multicolumn{1}{c|}{0.523}  & \multicolumn{1}{c|}{0.542}  & 0.550   \\ \hline
\end{tabular}
\end{table}

\begin{table*}[!hbp]
\caption{The MAP@50 performances of MIRFlickr-25K, NUS-WIDE and Wikipedia at various hashing code lengths}
\label{MAP_50}
\begin{tabular}{|c|c|ccc|ccc|ccc|}
\hline
\multirow{2}{*}{Task} & \multirow{2}{*}{Method} & \multicolumn{3}{c|}{NUSWIDE}                                       & \multicolumn{3}{c|}{MIRFlickr}                                     & \multicolumn{3}{c|}{Wikipedia}                                     \\ \cline{3-11} 
                      &                         & \multicolumn{1}{c|}{16-bit} & \multicolumn{1}{c|}{32-bit} & 64-bit & \multicolumn{1}{c|}{16-bit} & \multicolumn{1}{c|}{32-bit} & 64-bit & \multicolumn{1}{c|}{16-bit} & \multicolumn{1}{c|}{32-bit} & 64-bit \\ \hline
\multirow{4}{*}{I2T}  & DJSRH                   & \multicolumn{1}{c|}{0.724}  & \multicolumn{1}{c|}{0.773}  & 0.798  & \multicolumn{1}{c|}{0.810}   & \multicolumn{1}{c|}{0.843}  & 0.862  & \multicolumn{1}{c|}{0.388}  & \multicolumn{1}{c|}{0.403}  & 0.412  \\ \cline{2-11} 
                      & JDSH                    & \multicolumn{1}{c|}{0.736}  & \multicolumn{1}{c|}{0.793}  & 0.832  & \multicolumn{1}{c|}{0.832}  & \multicolumn{1}{c|}{0.853}  & 0.882  & \multicolumn{1}{c|}{—}      & \multicolumn{1}{c|}{—}      & —      \\ \cline{2-11} 
                      & DSAH                    & \multicolumn{1}{c|}{0.775}  & \multicolumn{1}{c|}{0.805}  & 0.818  & \multicolumn{1}{c|}{0.863}  & \multicolumn{1}{c|}{0.877}  & 0.895  & \multicolumn{1}{c|}{0.416}  & \multicolumn{1}{c|}{0.430}   & 0.438  \\ \cline{2-11} 
                      & ASSPH                   & \multicolumn{1}{c|}{0.810}   & \multicolumn{1}{c|}{0.845}  & 0.863  & \multicolumn{1}{c|}{0.887}  & \multicolumn{1}{c|}{0.902}  & 0.923  & \multicolumn{1}{c|}{0.477}  & \multicolumn{1}{c|}{0.467}  & 0.477  \\ \hline \hline
\multirow{4}{*}{T2I}  & DJSRH                   & \multicolumn{1}{c|}{0.712}  & \multicolumn{1}{c|}{0.744}  & 0.771  & \multicolumn{1}{c|}{0.786}  & \multicolumn{1}{c|}{0.822}  & 0.835  & \multicolumn{1}{c|}{0.611}  & \multicolumn{1}{c|}{0.635}  & 0.646  \\ \cline{2-11} 
                      & JDSH                    & \multicolumn{1}{c|}{0.721}  & \multicolumn{1}{c|}{0.785}  & 0.794  & \multicolumn{1}{c|}{0.825}  & \multicolumn{1}{c|}{0.864}  & 0.878  & \multicolumn{1}{c|}{—}      & \multicolumn{1}{c|}{—}      & —      \\ \cline{2-11} 
                      & DSAH                    & \multicolumn{1}{c|}{0.770}   & \multicolumn{1}{c|}{0.790}   & 0.804  & \multicolumn{1}{c|}{0.846}  & \multicolumn{1}{c|}{0.860}   & 0.881  & \multicolumn{1}{c|}{0.644}  & \multicolumn{1}{c|}{0.650}   & 0.660   \\ \cline{2-11} 
                      & ASSPH                   & \multicolumn{1}{c|}{0.776}  & \multicolumn{1}{c|}{0.797}  & 0.822  & \multicolumn{1}{c|}{0.854}  & \multicolumn{1}{c|}{0.878}  & 0.884  & \multicolumn{1}{c|}{0.634}  & \multicolumn{1}{c|}{0.628}  & 0.634  \\ \hline
\end{tabular}
\end{table*}

\section{MAP@50 Performance}

In Table~\ref{MAP_50}, we report the MAP@50 performance on NUSWIDE, MIRFlickr, and Wikipedia datasets with the same parameters as in MAP@all experiments.
Here, we still directly cite the experimental results reported in their original papers. We want to higlight that DJSRH, JDSH and DSAH all focus on optimizing MAP@50, as their original papers only report the results of MAP@50 but not MAP@all. However, even when considering MAP@50 metris, our model ASSPH still outperforms those three methods in almost all the cases, which clearly demonstrates that our model is not optimized for a specific metric and the improvement of our model is significant.
%

\end{document}